\documentclass[showpacs,showkeys,preprint]{revtex4-1}
\usepackage[latin9]{inputenc}
\usepackage{amsmath}
\usepackage{graphicx}

\makeatletter

\providecommand{\tabularnewline}{\\}

\makeatother

\begin{document}
\title{Probing Top Quark FCNC $tq\gamma$ and $tqZ$ Couplings at Future
Electron-Proton Colliders}
\author{O. Cakir}
\affiliation{Department of Physics, Ankara University, 06100, Ankara, Turkey}
\author{A. Yilmaz}
\affiliation{Department of Electrical and Electronics Engineering, Giresun University,
28200, Giresun, Turkey }
\author{I. Turk Cakir}
\affiliation{Department of Energy Systems Engineering, Giresun University, 28200,
Giresun, Turkey}
\author{A. Senol, H. Denizli}
\affiliation{Department of Physics, Bolu Abant Izzet Baysal University, 14280 Bolu,
Turkey }
\begin{abstract}
The top quark flavor changing neutral current (FCNC) processes are
extremely suppressed within the Standard Model (SM) of particle physics.
However, they could be enhanced in a new physics model Beyond the
Standard Model (BSM). The top quark FCNC interactions would be a good
test of new physics at present and future colliders. Within the framework
of the BSM models, these interactions can be described by an effective
Lagrangian. In this work, we study $tq\gamma$ and $tqZ$ effective
FCNC interaction vertices through the process $e^{-}p\rightarrow e^{-}Wq+X$
at future electron proton colliders, projected as Large Hadron electron
Collider (LHeC) and Future Circular Collider-hadron electron (FCC-he).
The cross sections for the signal have been calculated for different
values of parameters $\lambda_{q}$ for $tq\gamma$ vertices and $\kappa_{q}$
for $tqZ$ vertices. Taking into account the relevant background we
estimate the attainable range of signal parameters as a function of
the integrated luminosity and present contour plots of couplings for
different significance levels including detector simulation. 
\end{abstract}
\maketitle

\section{Introduction}

Within the framework of the Standard Model (SM) of particle physics,
the top quark (with mass $m_{t}\sim173$ GeV) being the heaviest of
fundamental fermions decays to a bottom quark and a $W$ boson (most
frequently) while it's decays to light down type quarks are suppressed
due to the Cabibbo-Kobayashi-Maskawa (CKM) matrix \cite{Cabibbo63,Kobayashi73}.
It is also known that flavor changing neutral current (FCNC) transitions
in the up-sector or down-sector are absent at tree level. However,
these transitions at the loop level are highly suppressed due to the
Glashow-Iliopoulos-Maiani (GIM) mechanism \cite{GIM70}. Branching
ratios are BR$(t\to c\gamma)\sim10^{-14}$, BR$(t\to cZ)\sim10^{-14}$,
BR$(t\to cg)\sim10^{-12}$ and BR$(t\to cH)\sim10^{-15}$, and the
branchings for top to up quark transitions are about one order smaller,
which are well beyond the current sensitivity of the Large Hadron
Collider (LHC) experiments. These decay modes could be enhanced in
some extensions of the SM, for instance due to the presence of new
virtual particles in the loops. Therefore, from both theoretical and
experimental perspective, studying the top quark FCNC interactions
is an important component of the top quark physics program.

The ATLAS and CMS experiments have significantly improved previous
exclusion limits on the top quark FCNC couplings. The experimental
$95\%$ confidence level (C.L.) upper limits on the branching fractions
of the top quark FCNC decays obtained at the LHC are summarized as
follows: BR$(t\to ug)\leq4.0\times10^{-5}$, BR$(t\to cg)\leq2.0\times10^{-4}$
\cite{ATLAS2016a}; BR$(t\to u\gamma)\leq1.3\times10^{-4}$, BR$(t\to c\gamma)\leq1.7\times10^{-3}$
\cite{CMS2016a}; BR$(t\to uH)\leq2.4\times10^{-3}$ and BR$(t\to cH)\leq2.2\times10^{-3}$
\cite{ATLAS15}. Recently, a combined result for the $tqZ$ couplings
(through anomalous $tZ$ production) has improved the limits BR$(t\to uZ)\leq2.2\times10^{-4}$
and BR$(t\to cZ)\leq4.9\times10^{-4}$ \cite{Sirunyan17}. At the
High Luminosity Large Hadron Collider (HL-LHC) with $L_{int}=3$ ab$^{-1}$
the limits on the top FCNC are estimated to be BR$(t\to q\gamma)\leq2.5\times10^{-5}$
\cite{ATLAS13}, BR$(t\to uZ)\leq4.3\times10^{-5}$ and BR$(t\to cZ)\leq5.8\times10^{-5}$
\cite{ATLAS16b} at $95\%$ C.L.

Phenomenologically, the sensitivities to the top quark FCNC interactions
have been estimated on the branching ratio BR$(t\to uZ/u\gamma)\simeq10^{-5}$
for the HL-LHC with $\sqrt{s}=14$ TeV and $L_{int}=3$ ab$^{-1}$
, and the branching ratio BR$(t\to uZ/u\gamma)\simeq10^{-6}$ for
Future Circular Collider-hadron hadron (FCC-hh) with $\sqrt{s}=100$
TeV and $L_{int}=10$ ab$^{-1}$ in Ref. \cite{Aguilar17}, while
the bounds have been estimated an order of magnitude larger for BR$(t\to cZ/c\gamma)$.

The future hadron electron collider projects currently under consideration
are the Large Hadron electron Collider (LHeC) \cite{LHeC} and Future
Circular Collider-hadron electron (FCC-he) \cite{FCC18}. The LHeC
comprises a 60 GeV electron beam that will collide with the 7 TeV
proton beam of LHC, having an integrated luminosity of $L_{int}=100$
fb$^{-1}$ per year, and planning to reach $1$ ab$^{-1}$ over the
years. On the other hand, the FCC-he mode is considered to be realized
by accelerating electrons up to 60 GeV and colliding them with the
proton beam at the energy of 50 TeV. A number of recent work exploring
the new physics capability and potential of the projected $ep$ colliders
have been reported in Refs. \cite{LHeC,FCC18,ICakir2017,Denizli17,Kumar17}.

In this work, we study the process $e^{-}p\rightarrow e^{-}Wq+X$
including $tq\gamma$ and $tqZ$ effective FCNC interaction vertices
at future hadron electron colliders, namely LHeC and FCC-he. The effective
Lagrangian is introduced and used in Section II to calculate the top
quark FCNC decay widths $\Gamma(t\to q\gamma)$ and $\Gamma(t\to qZ)$
and the branching ratios. The cross sections for the signal have been
calculated for different values of parameters $\lambda_{q}$ for $tq\gamma$
vertices and $\kappa_{q}$ for $tqZ$ vertices. We estimate the attainable
range of top quark FCNC parameters depending on the integrated luminosity
of the future ep colliders in section III. The signal and background
analysis including realistic detector effects have been performed,
and the contour plots of couplings $\kappa_{q}$ and $\lambda_{q}$
at different significance levels have been presented. Finally, we
summarize our results and conclude on the better limits for the top
FCNC branchings.

\section{Top Quark FCNC $tq\gamma$ and $tqZ$ Interactions}

At the electron-proton collision environment, top quark anomalous
FCNC interactions in the $tq\gamma$ and $tqZ$ vertices can be described
in a model independent effective Lagrangian 
\begin{align}
L_{eff} & =\frac{g_{e}}{2m_{t}}\bar{t}\sigma^{\mu\nu}(\lambda_{u}^{L}P_{L}+\lambda_{u}^{R}P_{R})uA_{\mu\nu}+\frac{g_{e}}{2m_{t}}\bar{t}\sigma^{\mu\nu}(\lambda_{c}^{L}P_{L}+\lambda_{c}^{R}P_{R})cA_{\mu\nu}\nonumber \\
+ & \frac{g_{W}}{4c_{W}m_{Z}}\bar{t}\sigma^{\mu\nu}(\kappa_{u}^{L}P_{L}+\kappa_{u}^{R}P_{R})uZ_{\mu\nu}+\frac{g_{W}}{4c_{W}m_{Z}}\bar{t}\sigma^{\mu\nu}(\kappa_{c}^{L}P_{L}+\kappa_{c}^{R}P_{R})cZ_{\mu\nu}+h.c.\label{eq:eq1}
\end{align}
where $g_{e}$ ($g_{W}$) is the electromagnetic (weak) coupling constant;
$c_{W}$ is the cosine of weak mixing angle; $\lambda_{q}^{L(R)}$
and $\kappa_{q}^{L(R)}$ are the strengths of anomalous top FCNC $tq\gamma$
and $tqZ$ couplings (where $q=u$,$c$), which vanish at the leading
order in the SM; $P_{L(R)}$ denotes the left (right) handed projection
operators. The photon field strength tensor is $A_{\mu\nu}$ and $Z$
boson field strenght tensor is $Z_{\mu\nu}$, and the anti-symmetric
tensor is $\sigma^{\mu\nu}=\frac{i}{2}[\gamma^{\mu},\gamma^{\nu}]$.
The effective Lagrangian is used to calculate both decay widths (for
the channels $t\to q\gamma$ and $t\rightarrow qZ$) and production
cross sections.

In addition to the usual decay channel $t\to W^{+}b$, the top quark
can also decay into up-type quarks ($u$ or $c$) associated with
a vector boson via FCNC as given in Eq. \ref{eq:eq1}. Considering
only the SM decay width and the FCNC interactions with electroweak
neutral gauge bosons, the top quark decay width ($\Gamma_{t}$) can
be written as

\begin{align}
\Gamma_{t} & =\Gamma(t\to W^{+}b)+\Gamma(t\to W^{+}s)+\Gamma(t\to W^{+}d)\nonumber \\
 & +\Gamma(t\to cZ)+\Gamma(t\to uZ)+\Gamma(t\to c\gamma)+\Gamma(t\to u\gamma)\label{eq:eq2}
\end{align}

\begin{figure}
\includegraphics[scale=0.8]{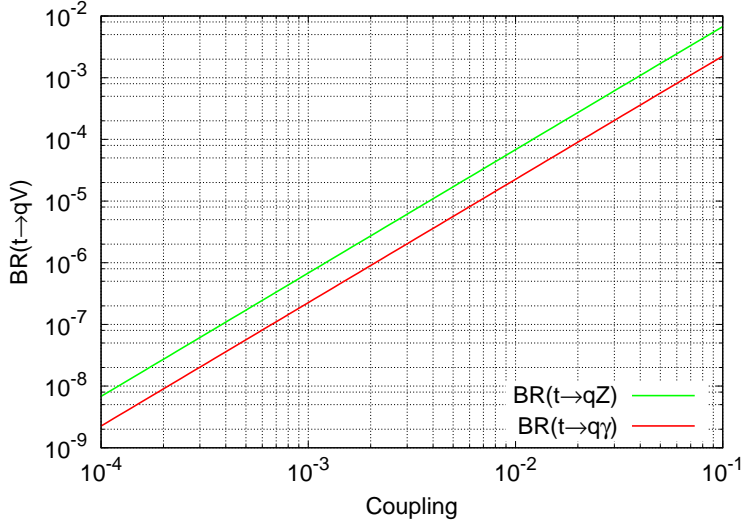}

\caption{Branching ratios for decay channels $t\to q\gamma$ and $t\to qZ$
depending on the FCNC coupling. \label{fig:fig1}}
\end{figure}

The dominant SM decay mode of top quark is $t\to W^{+}b$, the decay
width for this mode is given as

\begin{equation}
\Gamma(t\to W^{+}b)=\frac{\alpha|V_{tb}|^{2}}{16s_{W}^{2}}\frac{m_{t}^{3}}{m_{W}^{2}}(1-3m_{W}^{4}/m_{t}^{4}+2m_{W}^{6}/m_{t}^{6})\label{eq:eq3}
\end{equation}
at the leading order (LO), and it is improved to the next to leading
order (NLO) expression as given in Ref. \cite{Li91}. The ratios of
the SM decay widths are calculated as $\Gamma(t\to W^{+}s)/\Gamma(t\to W^{+}b)\simeq|V_{ts}|^{2}/|V_{tb}|^{2}\simeq1.495\times10^{-3}$
and $\Gamma(t\to W^{+}d)/\Gamma(t\to W^{+}b)\simeq|V_{td}|^{2}/|V_{tb}|^{2}\simeq6.318\times10^{-5}$
\cite{Tanabashi18}. The top quark FCNC partial decay widths are

\begin{equation}
\Gamma(t\to q\gamma)=\frac{\alpha}{4}(\lambda_{qL}^{2}+\lambda_{qR}^{2})m_{t}\label{eq:eq4}
\end{equation}
for the $t\to q\gamma$ channel, while the other partial decay widths
are

\begin{equation}
\Gamma(t\to qZ)=\frac{\alpha}{32s_{W}^{2}c_{W}^{2}m_{Z}^{2}}(\kappa_{qL}^{2}+\kappa_{qR}^{2})m_{t}^{3}(1-m_{Z}^{2}/m_{t}^{2})(2-m_{Z}^{2}/m_{t}^{2}-m_{Z}^{4}/m_{t}^{4})\label{eq:eq5}
\end{equation}
for the $t\to qZ$ channel, where $q=u,c$. The branching ratios for
$t\to q\gamma$ and $t\to qZ$ decay channels depending on the FCNC
$tq\gamma$ and $tqZ$ couplings are shown in Fig. \ref{fig:fig1}.

\section{Sensitivities at Future ep Colliders}

\begin{figure}
\includegraphics[scale=0.4]{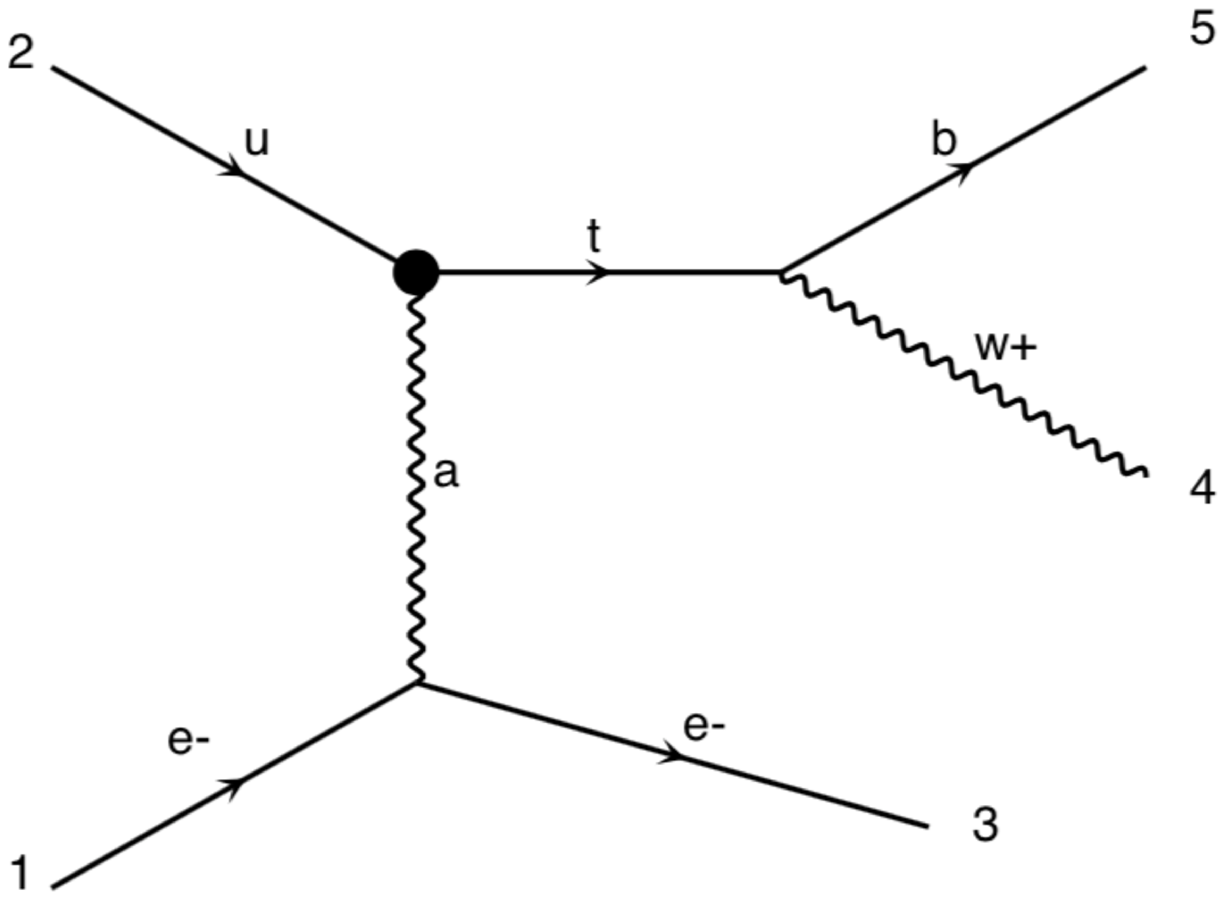}\includegraphics[scale=0.4]{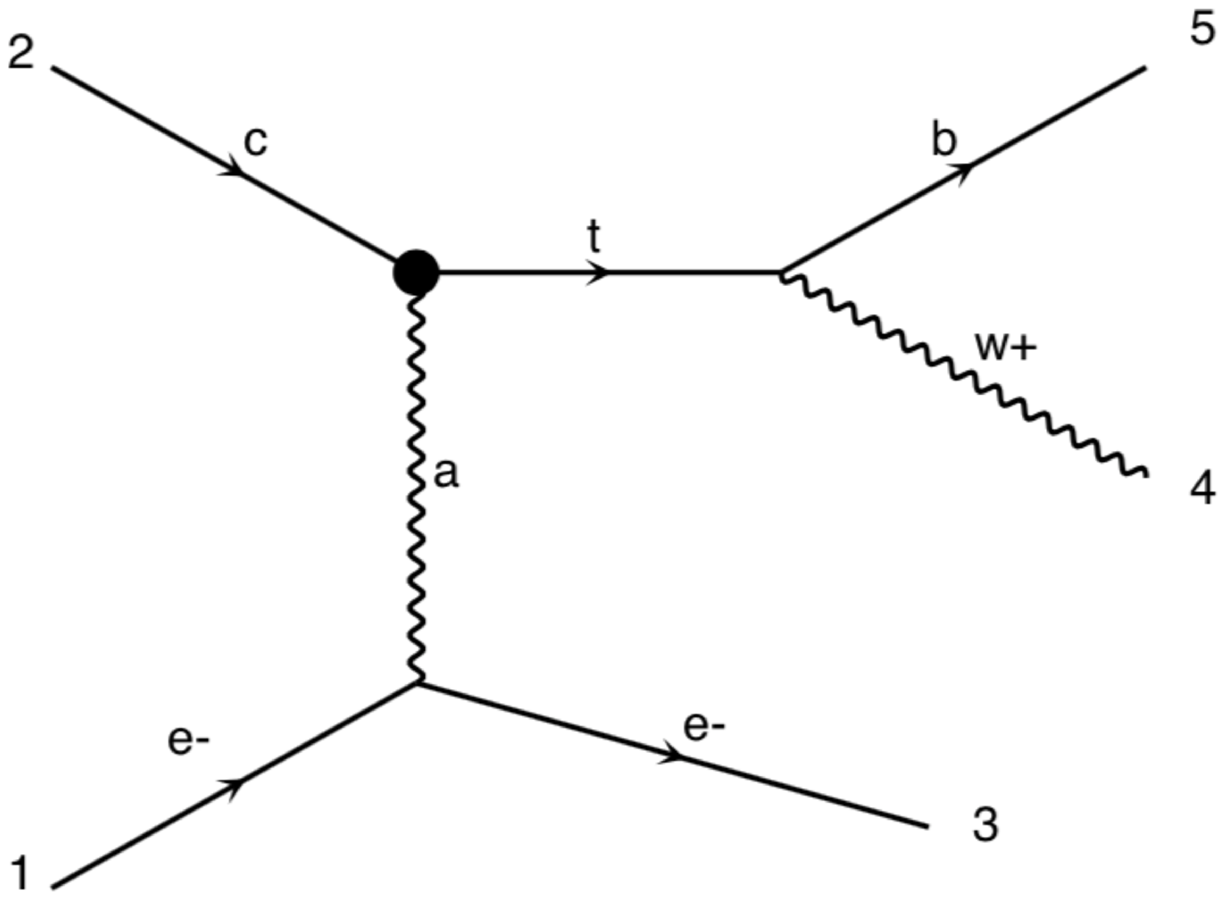}

\includegraphics[scale=0.4]{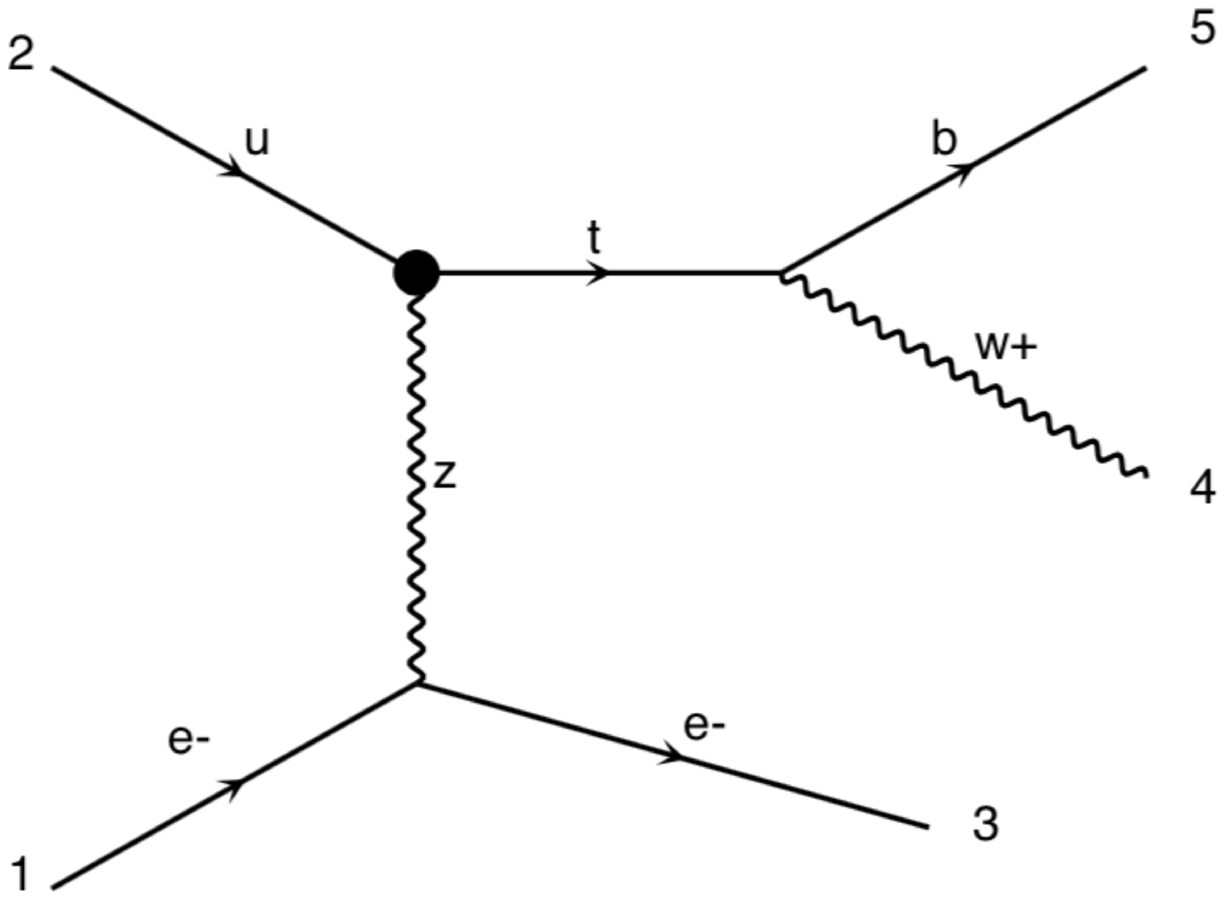}\includegraphics[scale=0.4]{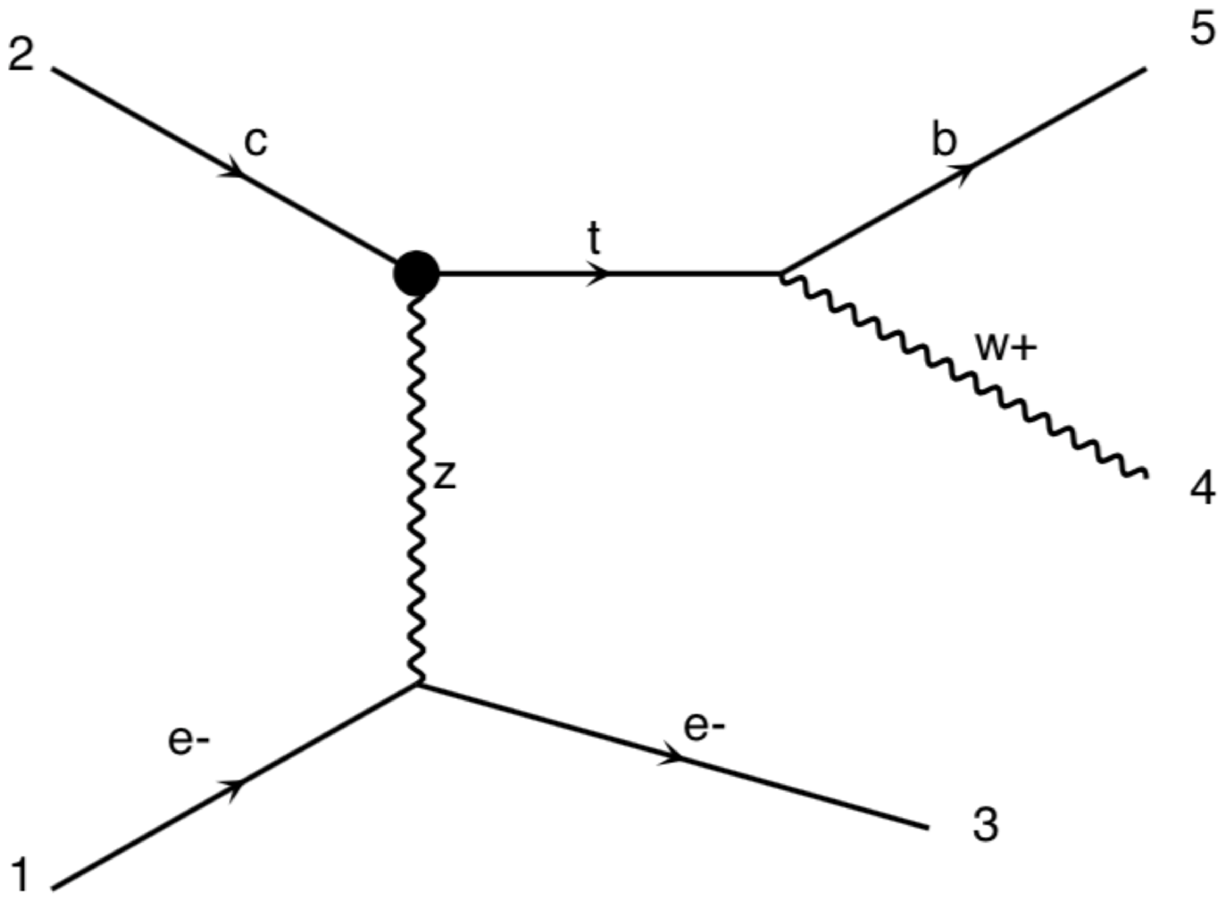}

\caption{Diagrams including top quark FCNC $tq\gamma$ and $tqZ$ interaction
vertices.\label{fig:fig2}}
\end{figure}

\begin{figure}[!hbt]
\includegraphics[scale=0.32]{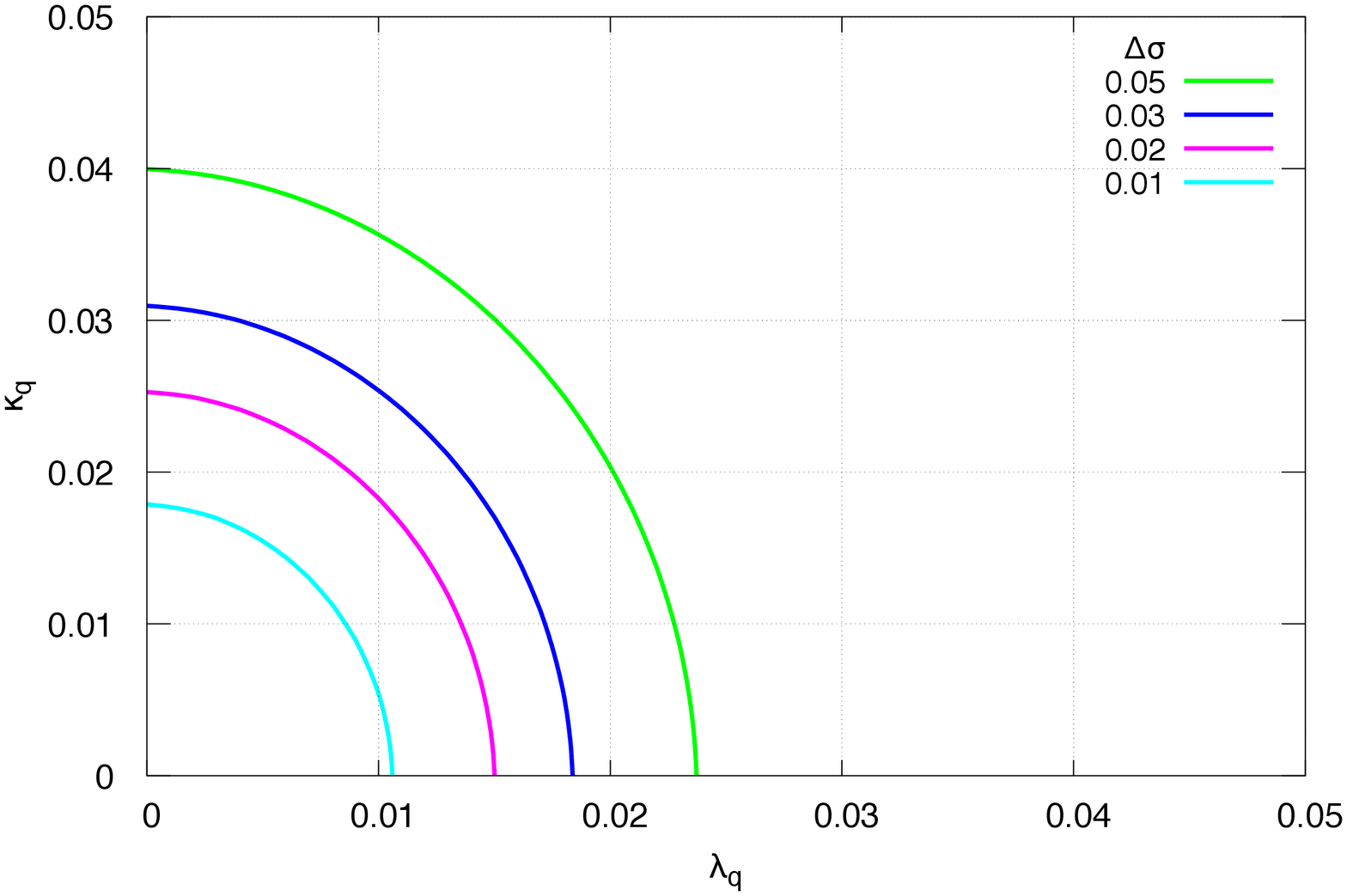} \includegraphics[scale=0.32]{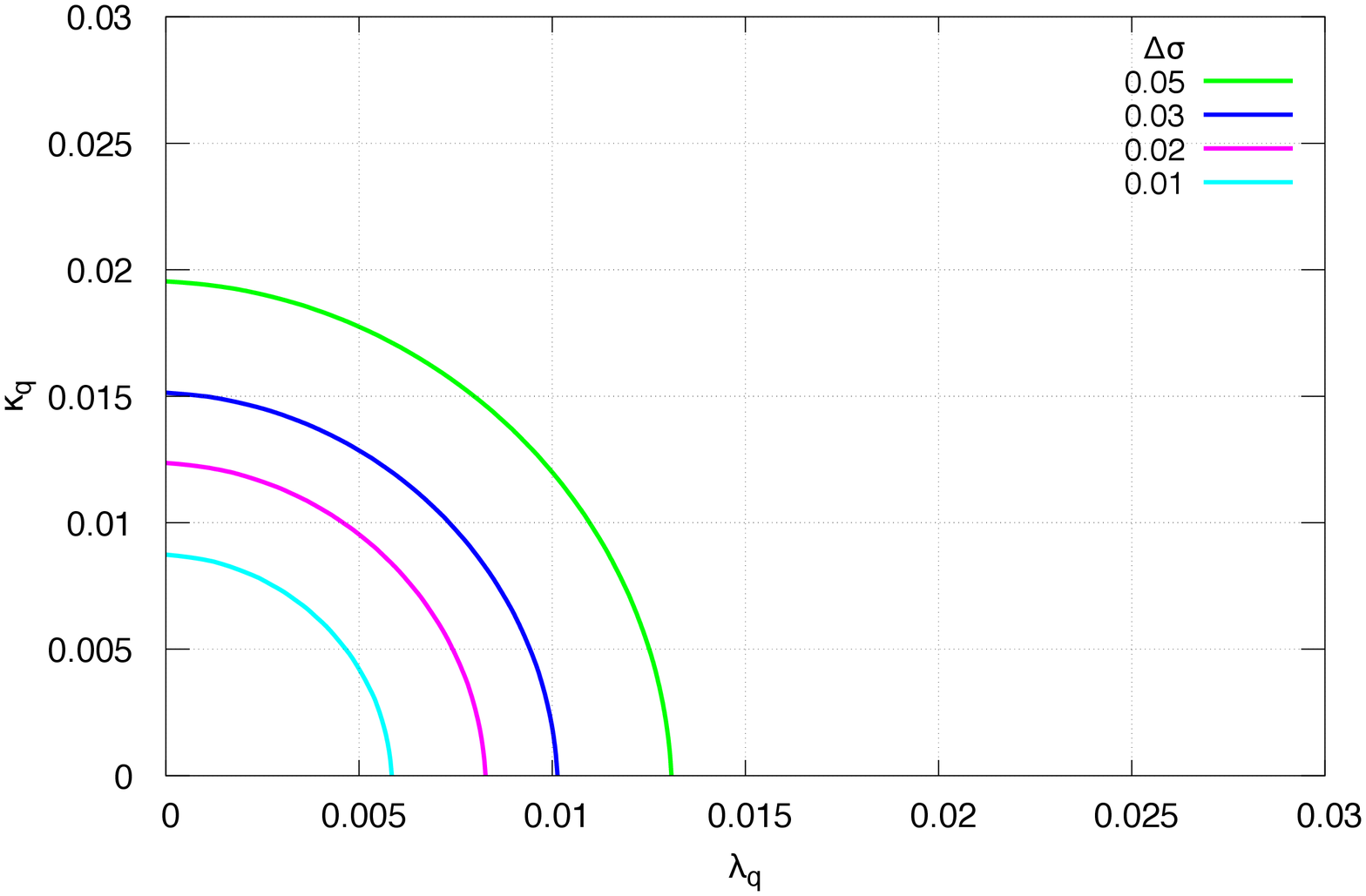}\caption{Contour plot for the top quark FCNC couplings $\kappa_{q}$ and $\lambda_{q}$
depending on the values (in pb) of signal cross sections at LHeC (left)
and FCC-he (right).\label{fig:fig3}}
\end{figure}

The production subprocess ($e^{-}q\to e^{-}W^{+}b$, where $q=u$,$c$)
including signal diagrams with $tq\gamma$ and $tqZ$ interaction
vertices is presented in Fig. \ref{fig:fig2}. The similar diagrams
for the subprocess ($e^{-}\bar{q}\to e^{-}W^{-}\bar{b}$) have also
been included in the calculation. The cross sections for the process
$e^{-}p\to e^{-}W^{\pm}q+X$ at different values of couplings $\kappa_{q}$
and $\lambda_{q}$ in the range of $(0.00-0.05)$ at LHeC and FCC-he
are given in Table \ref{tab:tab1}. The cross section increases when
the coupling parameters $\kappa_{q}$ and $\lambda_{q}$ grow in the
interested range. We plot the contours using Table \ref{tab:tab1}
to estimate the sensitivity to FCNC coupling parameters. The contour
lines correspond to different values of the signal cross sections
(where $\Delta\sigma$ denotes the signal cross section (in pb) when
the interfering background cross section is subtracted from the total
cross section) as shown in Fig. \ref{fig:fig3} for LHeC and FCC-he.
For a cross section value of the signal the sensitivity to coupling
parameter $\lambda_{q}$ is higher than the coupling parameter $\kappa_{q}$.

\begin{table}[!htp]
\caption{The cross section values (in pb) for process $e^{-}p\rightarrow e^{-}W^{\pm}q+X$
at LHeC depending on different values of the couplings. The numbers
in parenthesis denote the cross sections (in pb) at FCC-he. \label{tab:tab1}}

\centering{}%
\begin{tabular}{llccccccccc}
\hline 
Couplings &  & $\lambda_{q}$ = 0.00  &  & $\lambda_{q}$ = 0.01  &  & $\lambda_{q}$ = 0.02  &  & $\lambda_{q}$ = 0.03  &  & $\lambda_{q}$ = 0.05 \tabularnewline
\hline 
$\kappa_{q}$ = 0.00  &  & 2.3000 (8.6100) &  & 2.3094 (8.6421) &  & 2.3365 (8.7275) &  & 2.3805 (8.8737) &  & 2.5213 (9.3411)\tabularnewline
$\kappa_{q}$ = 0.01  &  & 2.3043 (8.6251) &  & 2.3136 (8.6574) &  & 2.3236 (8.7445) &  & 2.3852 (8.8914) &  & 2.5268 (9.3636)\tabularnewline
$\kappa_{q}$ = 0.02  &  & 2.3135 (8.6646) &  & 2.3406 (8.6956) &  & 2.3505 (8.7899) &  & 2.3957 (8.9344) &  & 2.5387 (9.4088)\tabularnewline
$\kappa_{q}$ = 0.03  &  & 2.3286 (8.7324) &  & 2.3390(8.7659) &  & 2.3666 (8.8518) &  & 2.4123 (9.0031) &  & 2.5559 (9.4776)\tabularnewline
$\kappa_{q}$ = 0.05  &  & 2.3782 (8.9341) &  & 2.3885 (8.9725) &  & 2.4173 (9.0690) &  & 2.4639 (9.2270) &  & 2.6082 (9.7070)\tabularnewline
\hline 
\end{tabular}
\end{table}

The process $e^{-}p\to e^{-}W^{\pm}q+X$ includes both the signal
and the background interfering with the signal. We calculate the cross
sections for this process to normalize the distributions from the
signal and background events. We take into account the main background
(B1: $e^{-}W^{\pm}q$) and include other background (B2: $e^{-}Zq$)
which contain at least three jets and one electron in the final state.
Here, QCD multijet backgrounds are not included in the analysis of
top quark FCNC $tq\gamma$ and $tqZ$ interactions.

In our calculations, we produce signal and background events by using
MadGraph 5\_aMC@NLO \citep{MadGrapH2}, with an effective Lagrangian
implementation through FeynRules \citep{FeynRules3} for the signal.
Afterwards the parton showering and detector fast simulations are
carried out with Pythia 6 \citep{Pythia4} and Delphes 3.4 \citep{Delphes5},
respectively.

\begin{figure}[!hbt]
\includegraphics[width=0.48\textwidth]{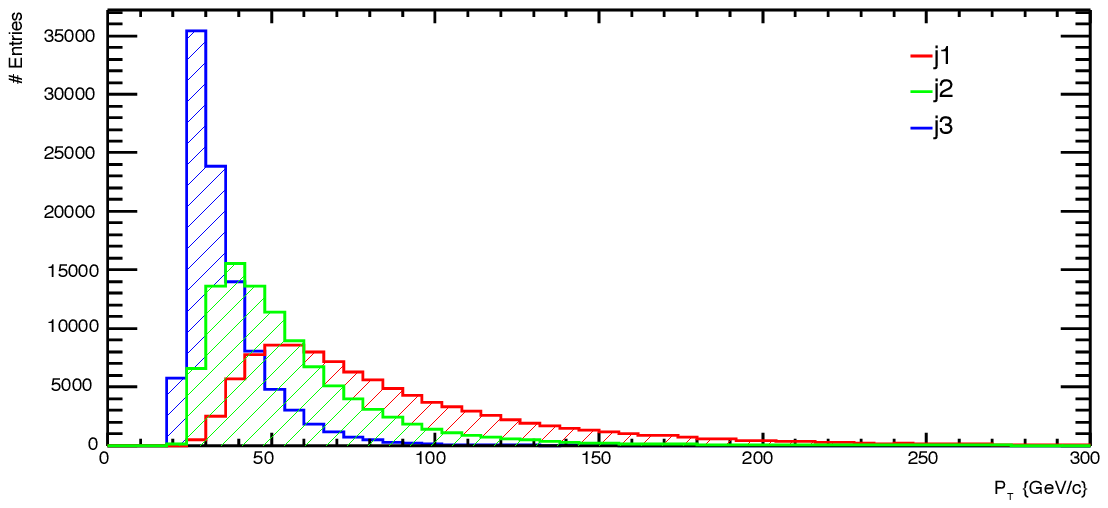}\includegraphics[width=0.49\textwidth]{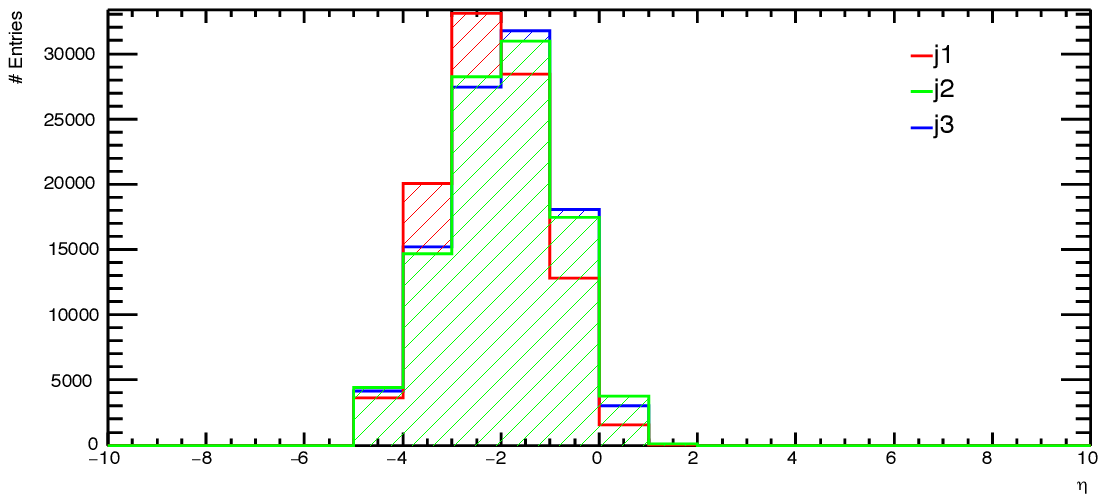}

\includegraphics[width=0.48\textwidth]{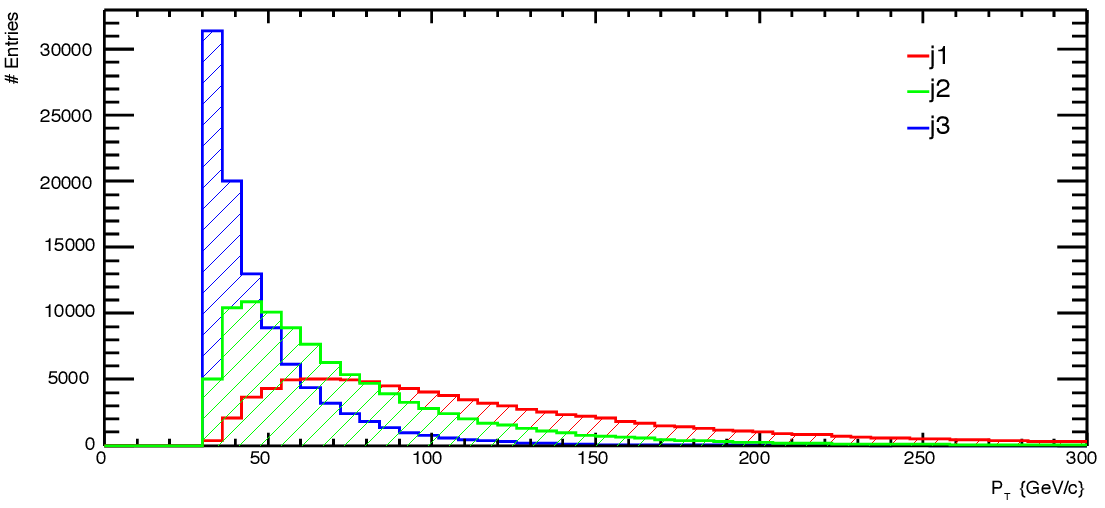}\includegraphics[width=0.49\textwidth]{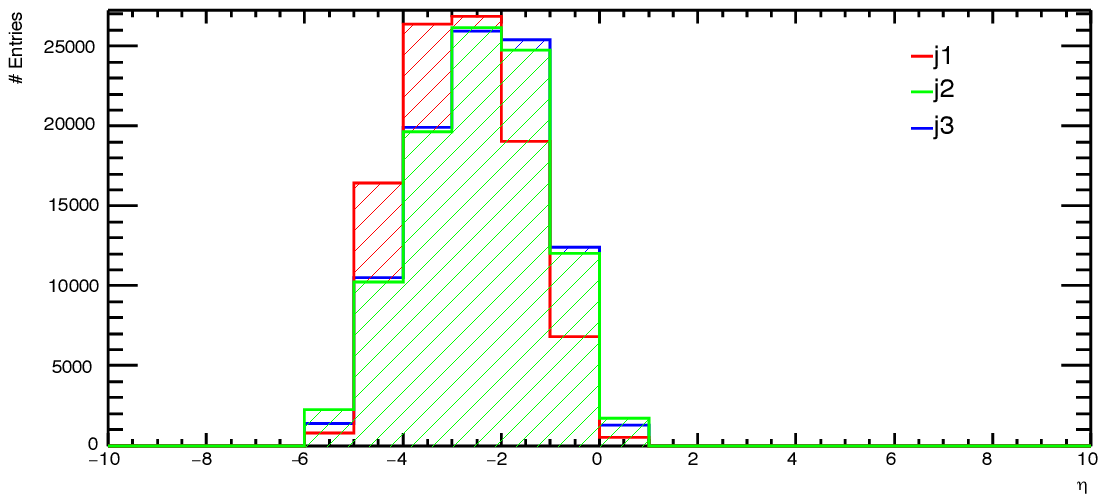} 

\caption{Transverse momentum ($p_{T}$) and pseudo-rapidity ($\eta$) distributions
of three jets from the process $e^{-}p\rightarrow e^{-}W^{\pm}q+X$
which includes both the interfering background and signal for $\kappa_{q}$
= $\lambda_{q}$ = 0.05 at LHeC (first row) and FCC-he (second row).\label{fig:fig4}}
\end{figure}

\begin{figure}[!hbt]
\includegraphics[width=0.48\textwidth]{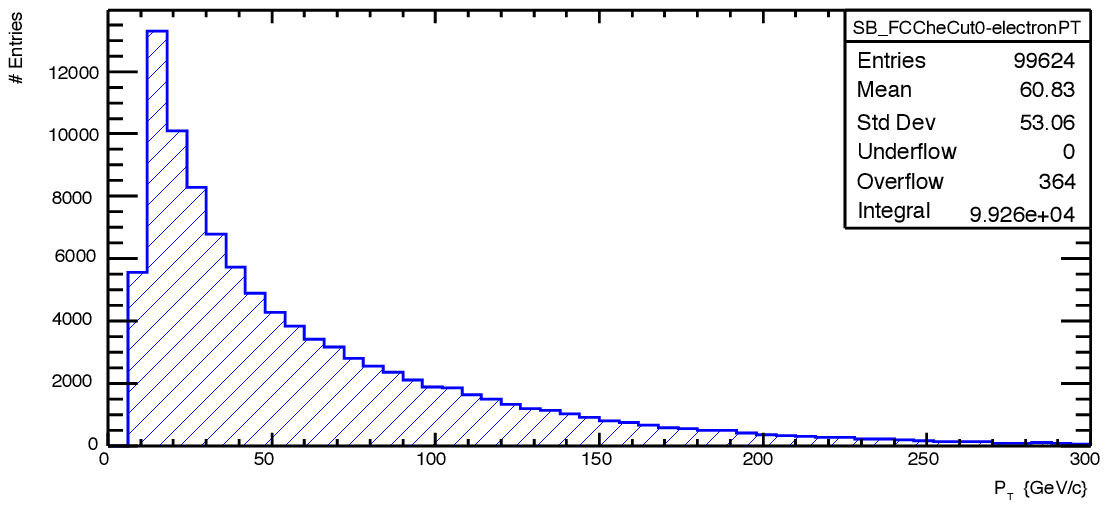} \includegraphics[width=0.49\textwidth]{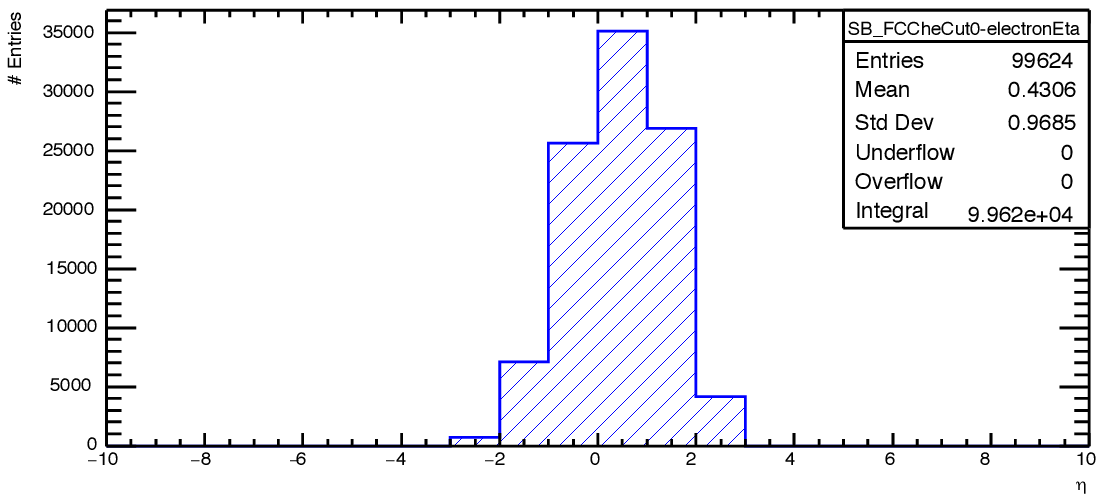}

\includegraphics[width=0.48\textwidth]{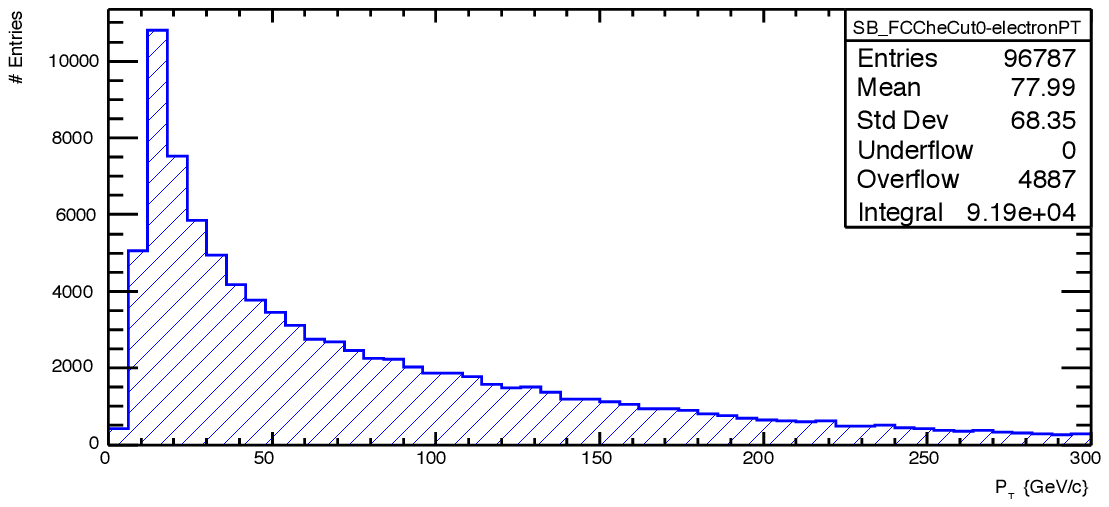}\includegraphics[width=0.49\textwidth]{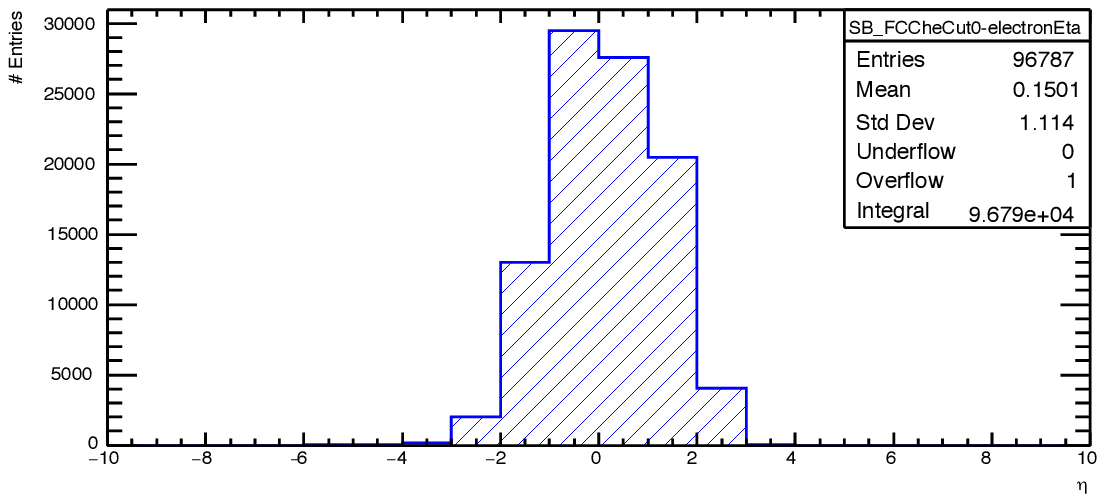}

\caption{Transverse momentum and pseudo-rapidity distributions of electron
from the process $e^{-}p\rightarrow e^{-}W^{\pm}q+X$ which includes
both the interfering background and signal for $\kappa_{q}$ = $\lambda_{q}$
= 0.05 at LHeC (first row) and FCC-he (second row). \label{fig:fig5}}
\end{figure}

The kinematical distributions for signal and interfering background
are given in Fig. \ref{fig:fig4} for LHeC and FCC-he. The transverse
momentum ($p_{T}$) (on the left) and rapidty ($\eta$) (on the right)
distributions of the leading jet, second leading jet and third leading
jet are presented in these figures. These distributions are obtained
after preselection of the events. For the analysis of signal and background
events, we also apply analysis cuts after the generator level pre-selection.
In order to select signal events we require having one electron and
three jets ordered according to the highest transverse momentum $p_{T}$.
Since there is an energy asymmetry in the electron-proton collisions,
the jets from the process mainly peaks in the backward region, hence
the pseudo-rapidy range for jets is taken as $-4<\eta<0$ in the analysis.
The transverse momentum $p_{T}$ and pseudo-rapidty $\eta$ distributions
of signal and main background have quite similar behaviour since we
deal with small couplings for the signal and we take into account
the interference of signal and background as well. In Fig. \ref{fig:fig5},
the kinematical distributions ($p_{T}$ and $\eta$) of electron in
the events are depicted. One of the specific aspects of the signal
is the occurrence of the high $p_{T}$ electron in the central $\eta$
region.

In the analysis, we require at least three jets and one electron in
the events, one of the jets should be $b$-tagged with leading jet
$p_{T}(j)>40$ GeV and other jets having $p_{T}(j)>30$ GeV and $\mid\eta(j)\mid<2.5$,
the electron with $p_{T}(e)>20$ GeV and $\mid\eta(e)\mid<2.5$ as
the cut flow given in Table \ref{tab:tab2}. Further steps in the
cut flow table include invariant mass intervals for selecting events
for the analysis.

\begin{table}[!htp]
\caption{Preselection and a set of cuts for the analysis of signal and background
events. \label{tab:tab2}}

\centering{}%
\begin{tabular}{lllll}
\hline 
Cuts  &  &  &  & Definition \tabularnewline
\hline 
Cut-0  &  &  &  & Preselection: $N_{jets}>=$ 3 and $N_{e}>=$ 1 \tabularnewline
Cut-1  &  &  &  & $b-$tag: one $b-$tagged jet ($j_{b}$) \tabularnewline
Cut-2  &  &  &  & Transverse momentum: $p_{T}(j_{2},j_{3})>$ 30 GeV and $p_{T}(j_{b})>$
40 GeV and $p_{T}(e)>$ 20 GeV \tabularnewline
Cut-3  &  &  &  & Pseudo-rapidity: -4 $<\eta(j_{b},j_{2},j_{3})<0$ and $|\eta(e)|<$
2.5 \tabularnewline
Cut-4  &  &  &  & $W$ boson mass: 50 $<M_{inv}^{rec}(j_{2},j_{3})<$ 100 GeV \tabularnewline
Cut-5  &  &  &  & Top quark mass: 130 $<M_{inv}^{rec}(j_{b},j_{2},j_{3})<$ 200 GeV \tabularnewline
\hline 
\end{tabular}
\end{table}

\begin{figure}[!hbt]
\includegraphics[scale=0.32]{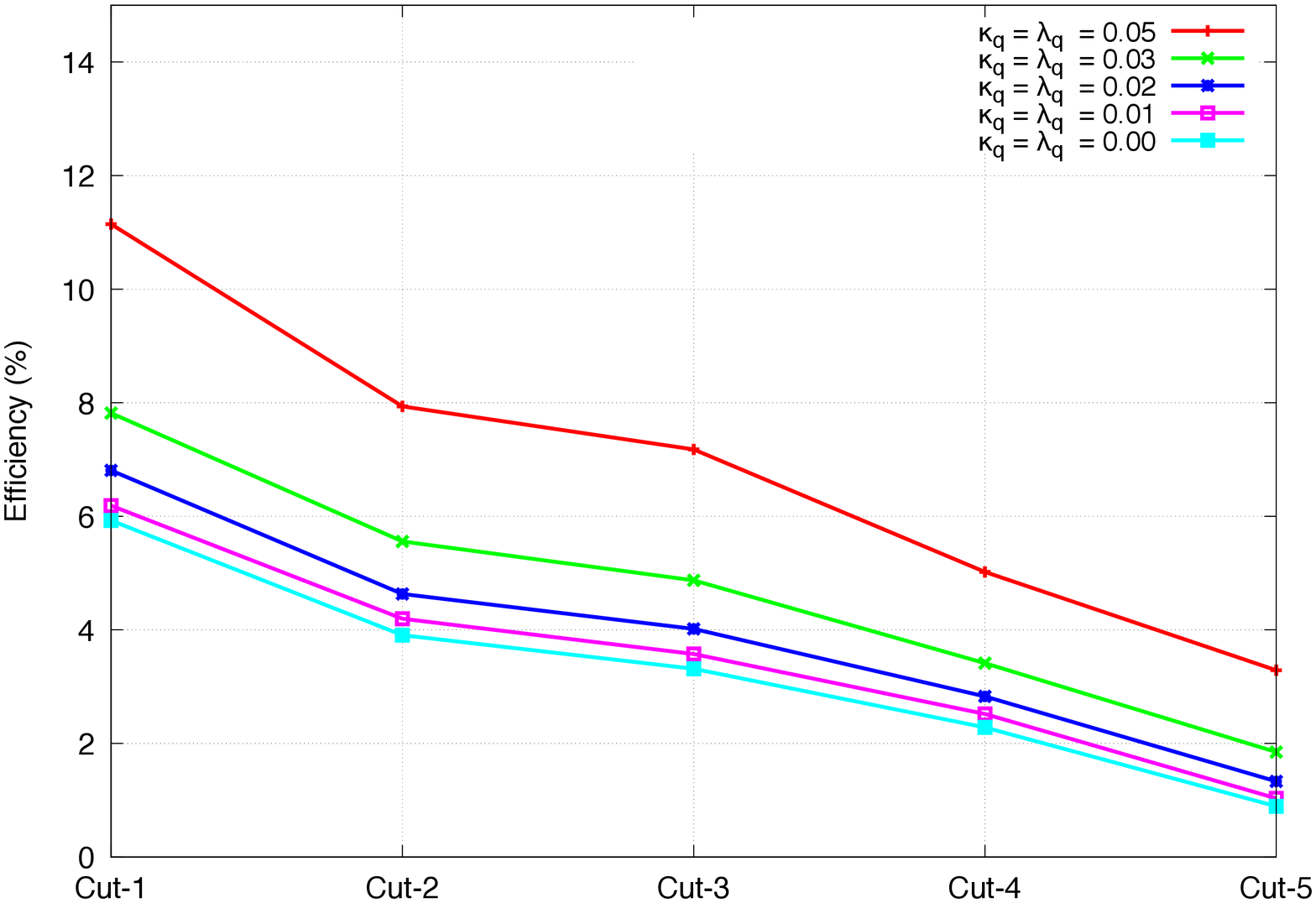} \includegraphics[scale=0.32]{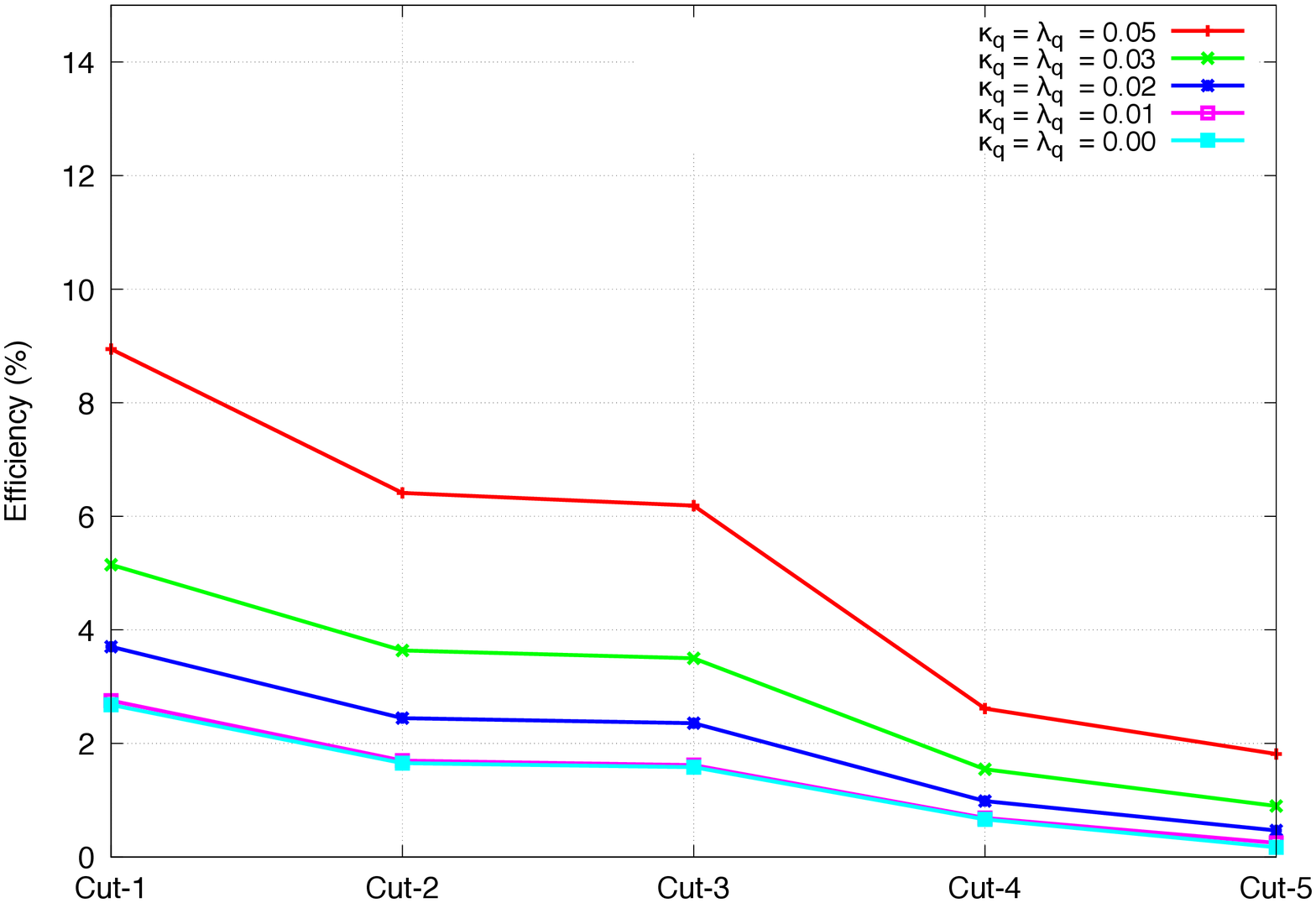}\caption{Efficiency plot for the cuts applied at each step for the analysis
of signal and background at LHeC (left) and FCC-he (right). \label{fig:fig6}}
\end{figure}

The cut efficiencies have been calculated after pre-selection for
signal and background as shown in Fig. \ref{fig:fig6} for LHeC and
FCC-he. We have larger cut efficiencies for higher values of the FCNC
couplings. Fig. \ref{fig:fig6} shows that the cut efficiency for
the background changes from $6\%$ to $1\%$ for Cut-1 to Cut-5, whereas
the cut efficiencies for the signal decrease from $11\%$ to $3.2\%$
for couplings $\kappa_{q}=\lambda_{q}=0.05$.

After Cut-5, the number of events for background and signal (different
values of couplings $\kappa_{q}$ and $\lambda_{q}$) are given in
Table \ref{tab:tab3} for LHeC and for FCC-he with an integrated luminosity
of $100$ fb$^{-1}$. For the coupling parameters $\kappa_{q}=\lambda_{q}=0.05$
we obtain the number of events 2153 (2844), while the background events
are 508 (231) at LHeC (FCC-he). Thus, the signal gives an enhancement
factor of 3.24 over the background for $\kappa_{q}=\lambda_{q}=0.05$,
whereas this factor is 0.17 for $\kappa_{q}=\lambda_{q}=0.01$. For
each cut step the number of events can be obtained from Table \ref{tab:tab3}
with the relative cut efficiency factors from Fig. \ref{fig:fig6}.

\begin{table}[!htp]
\caption{The number of events for main background (where $\kappa_{q}=0$ and
$\lambda_{q}=0$ ) and signal (where $\kappa_{q}\protect\neq0$ and
$\lambda_{q}\protect\neq0$) with different FCNC couplings $\kappa_{q}$
and $\lambda_{q}$ at LHeC and FCC-he with \textit{L}$_{int}=$ 100
fb$^{-1}$. The numbers in parenthesis denote the number of events
at FCC-he.\label{tab:tab3} }

\centering{}%
\begin{tabular}{llccccccccc}
\hline 
Couplings &  & $\lambda_{q}$ = 0.00  &  & $\lambda_{q}$ = 0.01  &  & $\lambda_{q}$ = 0.02  &  & $\lambda_{q}$ = 0.03 &  & $\lambda_{q}$ = 0.05 \tabularnewline
\hline 
$\kappa_{q}$ = 0.00  &  & 508 (231) &  & 558 (269) &  & 670 (462) &  & 894 (809) &  & 1469 (1765)\tabularnewline
$\kappa_{q}$ = 0.01  &  & 549 (259) &  & 595 (334) &  & 741 (491) &  & 901 (834) &  & 1624 (1818)\tabularnewline
$\kappa_{q}$ = 0.02  &  & 622 (421) &  & 646 (466) &  & 779 (647) &  & 971 (998) &  & 1633 (1932)\tabularnewline
$\kappa_{q}$ = 0.03  &  & 721 (576) &  & 765 (703) &  & 834 (915) &  & 1113 (1286) &  & 1841 (2227)\tabularnewline
$\kappa_{q}$ = 0.05  &  & 1037 (1292) &  & 1120 (1407) &  & 1256 (1652) &  & 1514 (1921) &  & 2153 (2844)\tabularnewline
\hline 
\end{tabular}
\end{table}

\begin{figure}[!hbt]
\begin{tabular}{cc}
\includegraphics[scale=0.3]{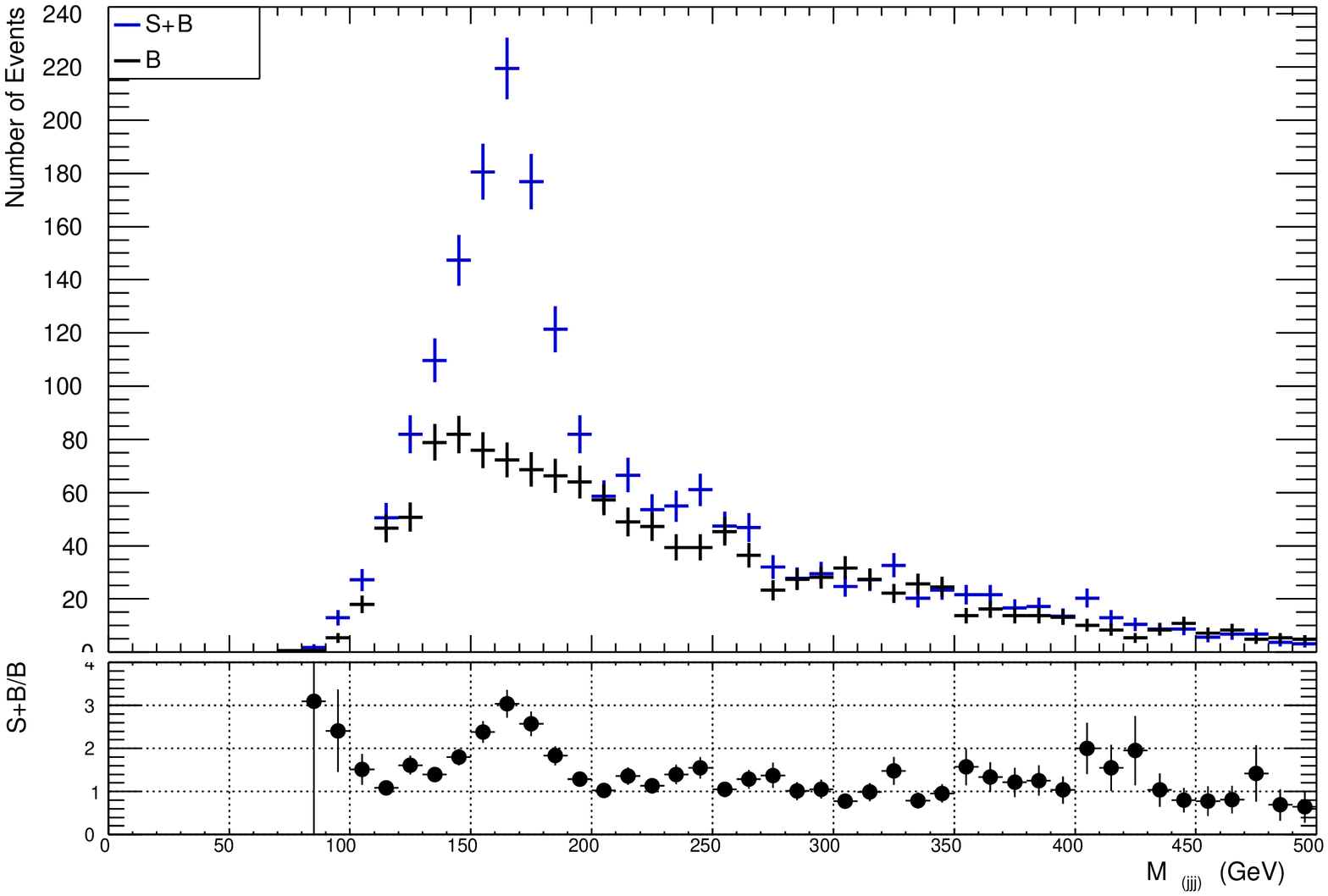} & \includegraphics[bb=0bp -65bp 653bp 446bp,scale=0.3]{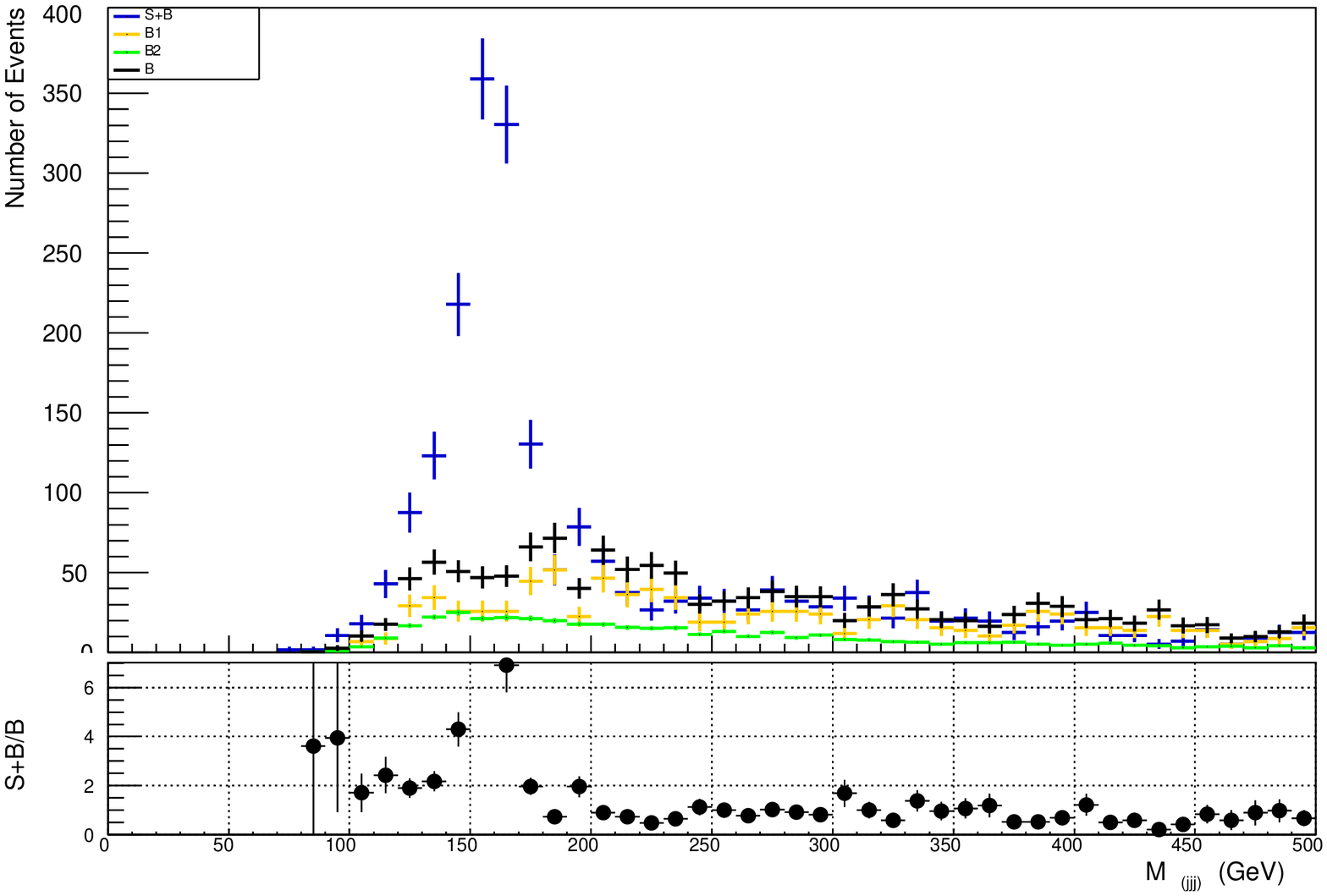}\tabularnewline
\includegraphics[scale=0.3]{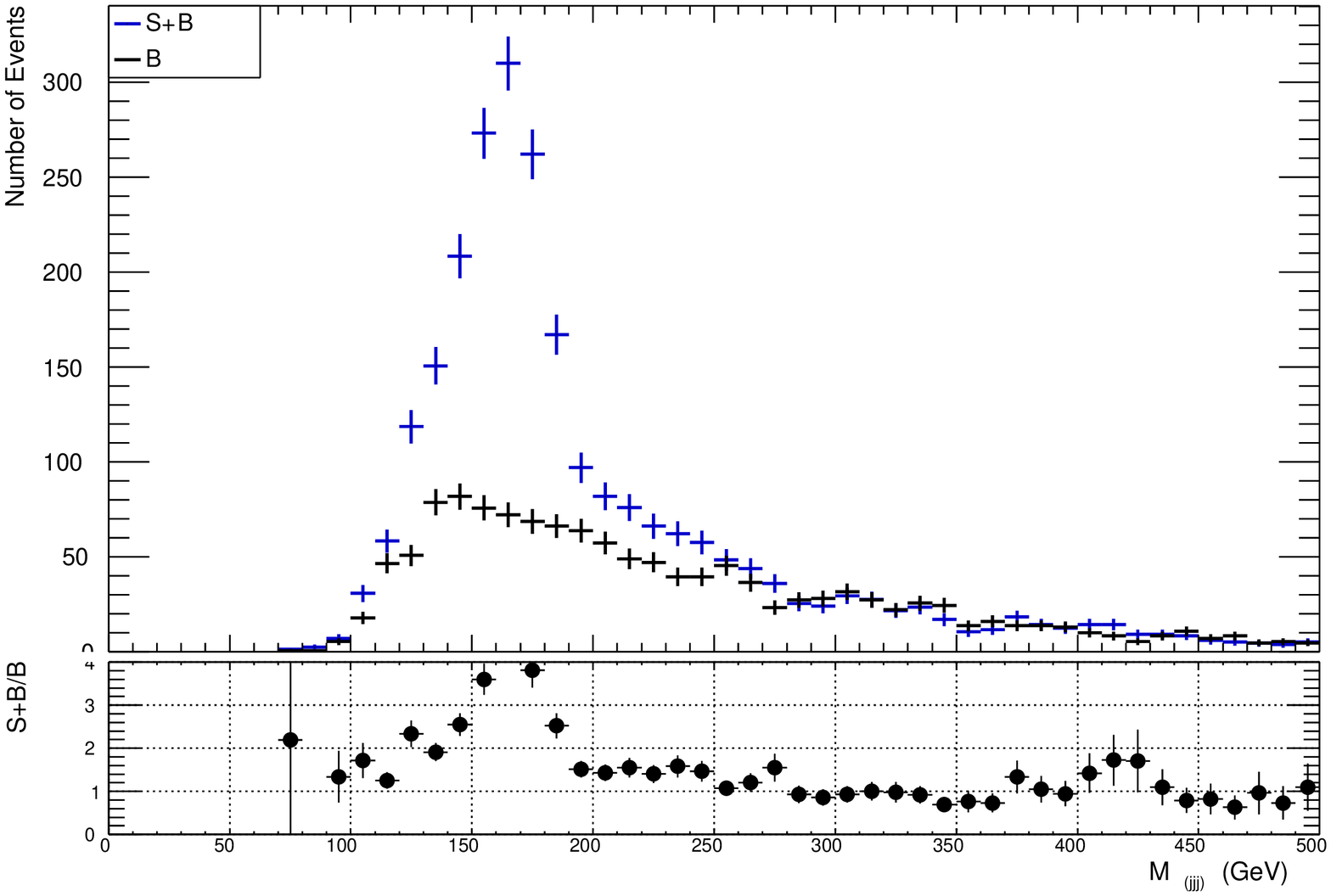} & \includegraphics[bb=0bp -65bp 655bp 442bp,scale=0.3]{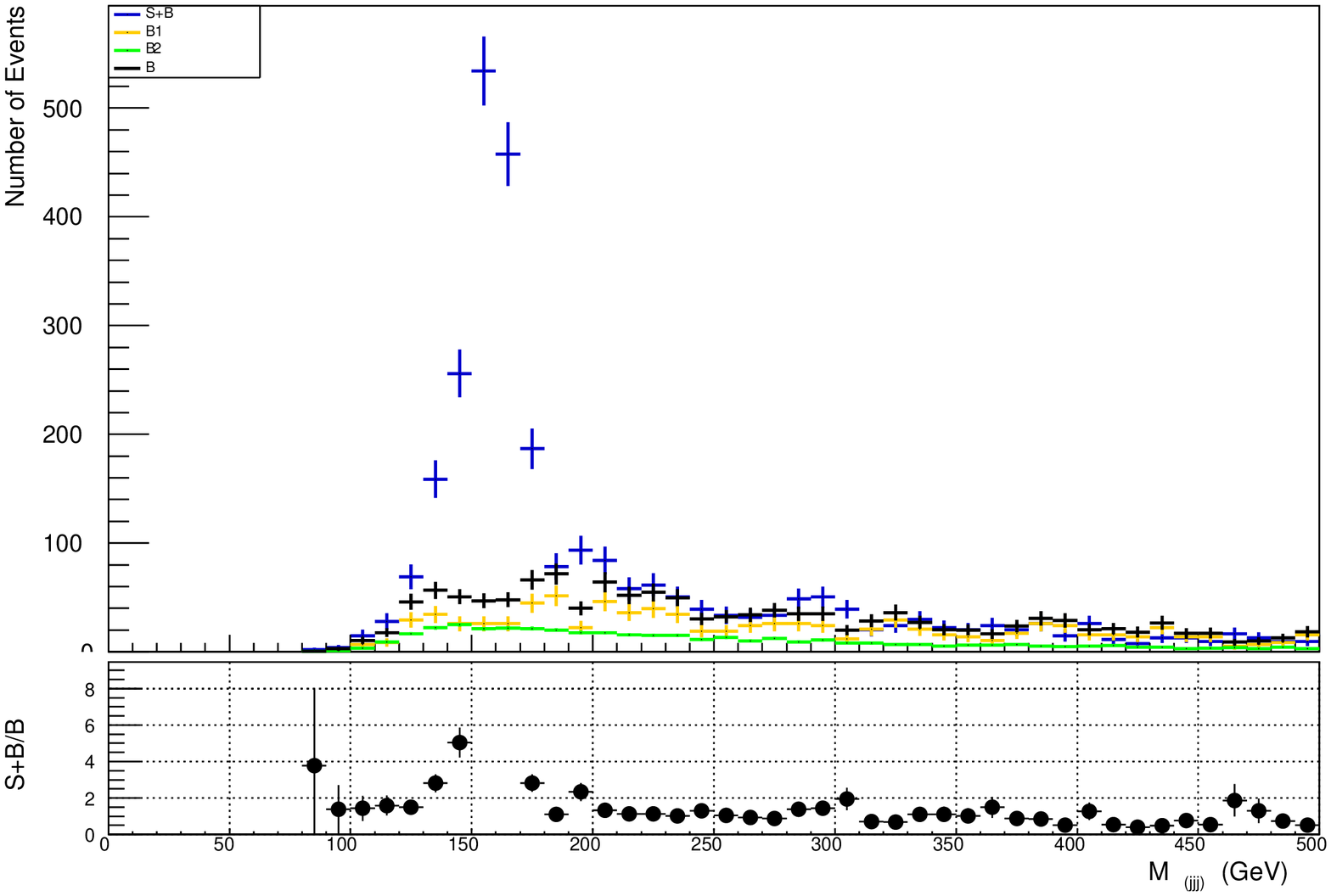}\tabularnewline
\includegraphics[scale=0.3]{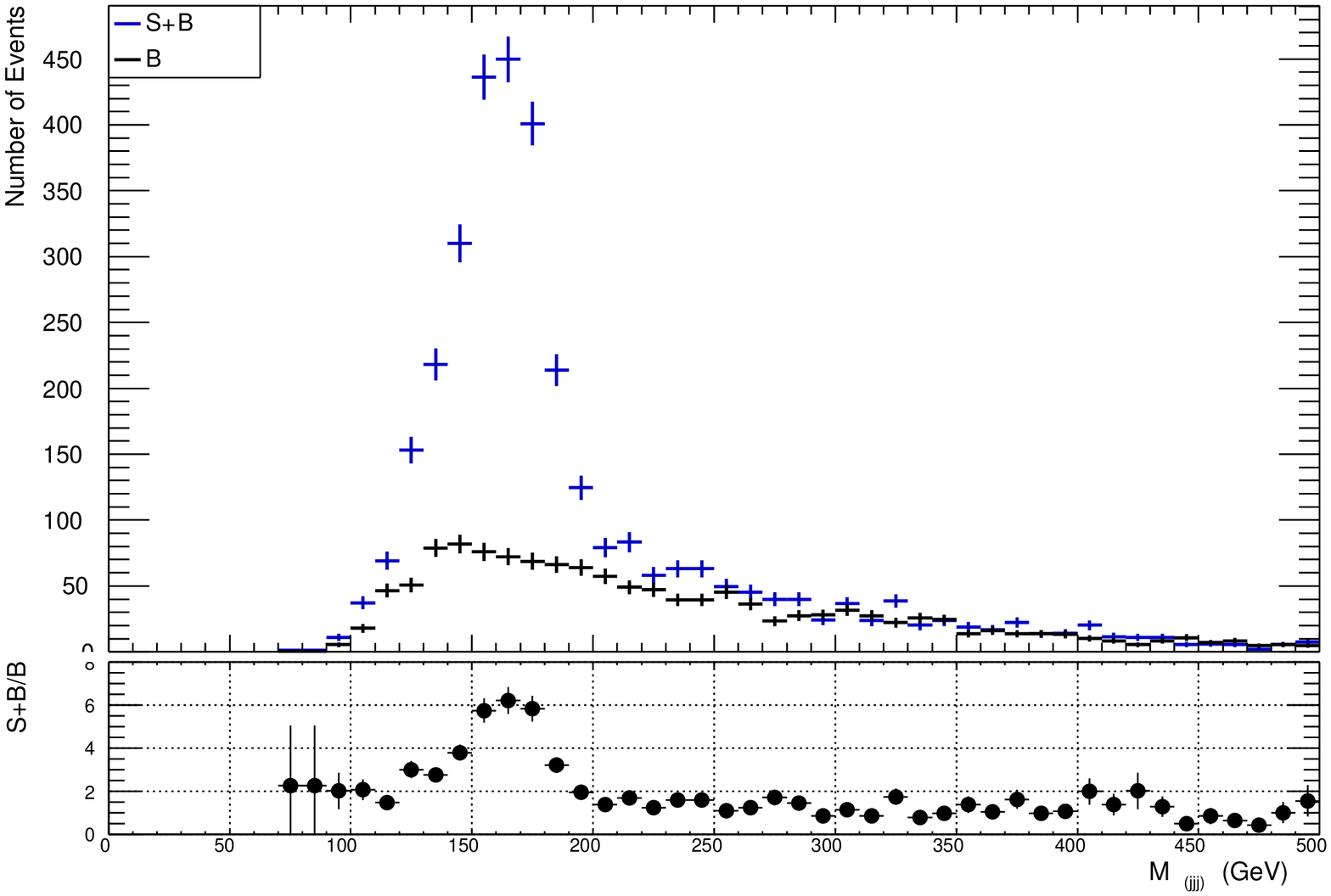} & \includegraphics[bb=0bp -65bp 652bp 440bp,scale=0.3]{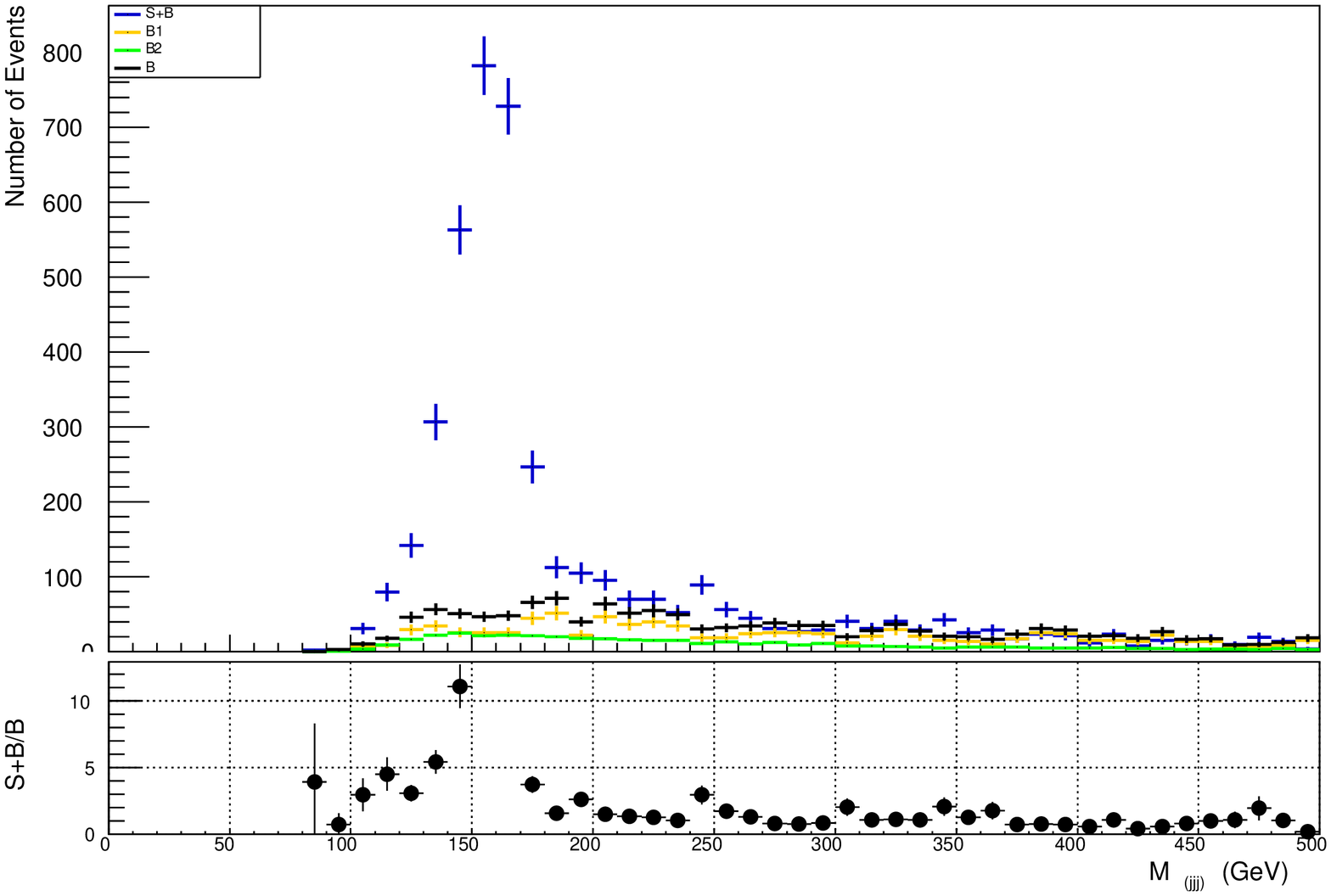}\tabularnewline
\end{tabular}

\caption{Invariant mass distributions of three jets (one of the jets is required
as \textit{b}-jet) for the signal + background (S+B) where B is the
main background at LHeC (first column) and FCC-he (second column).
First plot is for $\lambda_{q}=0,\:\kappa_{q}=0.05$ , second plot
is for $\lambda_{q}=0.05,\:\kappa_{q}=0$ and third plot is for $\lambda_{q}=\kappa_{q}=0.05$.
\label{fig:fig7}}
\end{figure}

We plot the invariant mass distribution of top quark reconstructed
from three jets (one of them is $b-$tagged) for different coupling
scenarios (at first row) $\lambda_{q}=0.0,\kappa_{q}=0.05$, (second
row) $\lambda_{q}=0.05,\kappa_{q}=0.0$ and (third row) $\lambda_{q}=0.05,\kappa_{q}=0.05$
as shown in Fig. \ref{fig:fig7} for LHeC and FCC-he. The ratio of
the $S+B$ and $B$ is more enhanced at top mass for equal coupling
scenario (c) when it is compared with the other scenarios (a) and
(b) as seen from Fig. \ref{fig:fig7}.

In order to quantify statistical significance ($SS$), we calculate
signal ($S$) and background ($B$) events after final cut. Here the
$SS$ is defined by

\begin{equation}
SS=\sqrt{2[(S+B)\ln(1+\frac{S}{B})-S]}\label{eq:eq6}
\end{equation}

\begin{figure}[!hbt]
\includegraphics[scale=0.3]{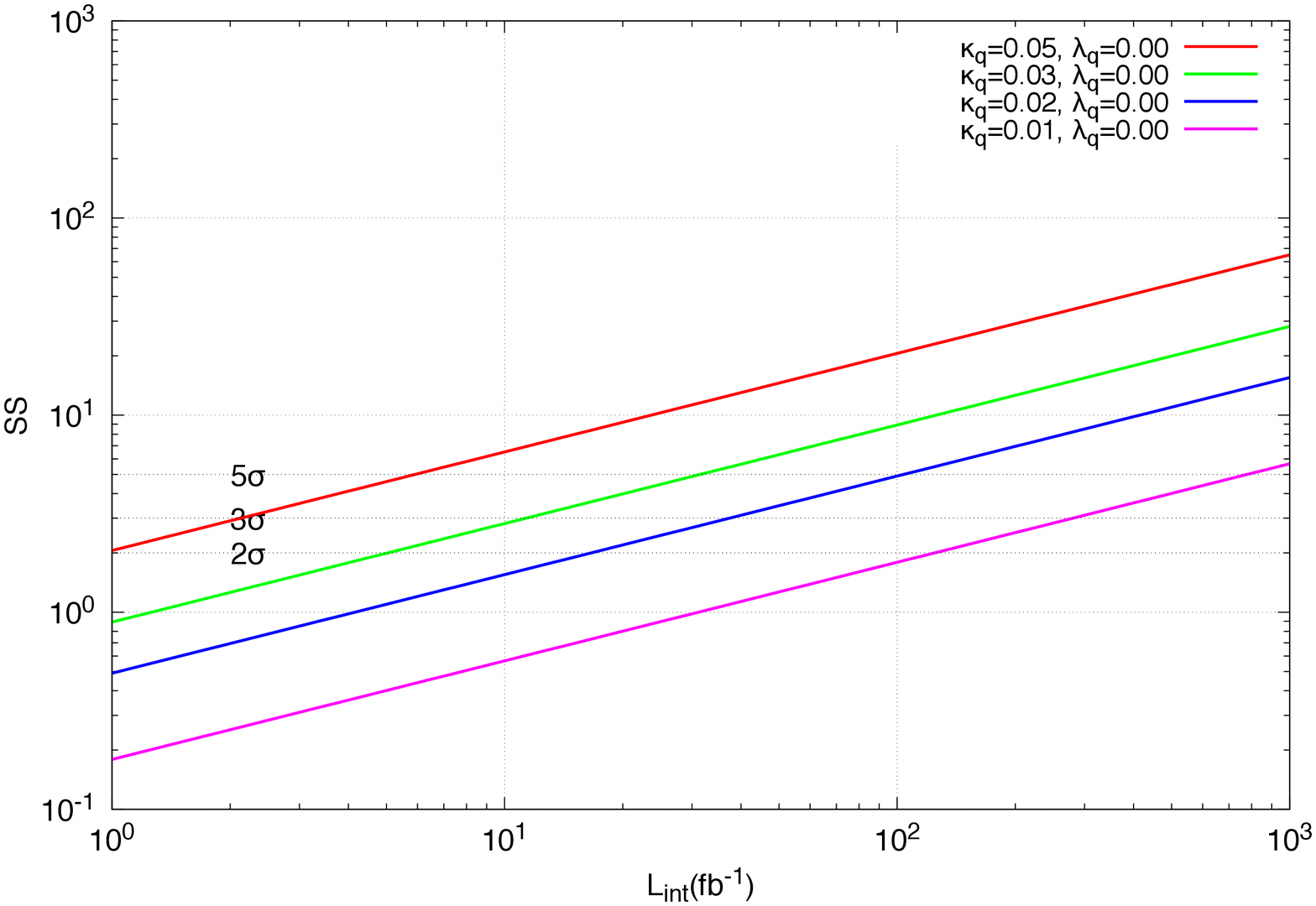}\includegraphics[scale=0.3]{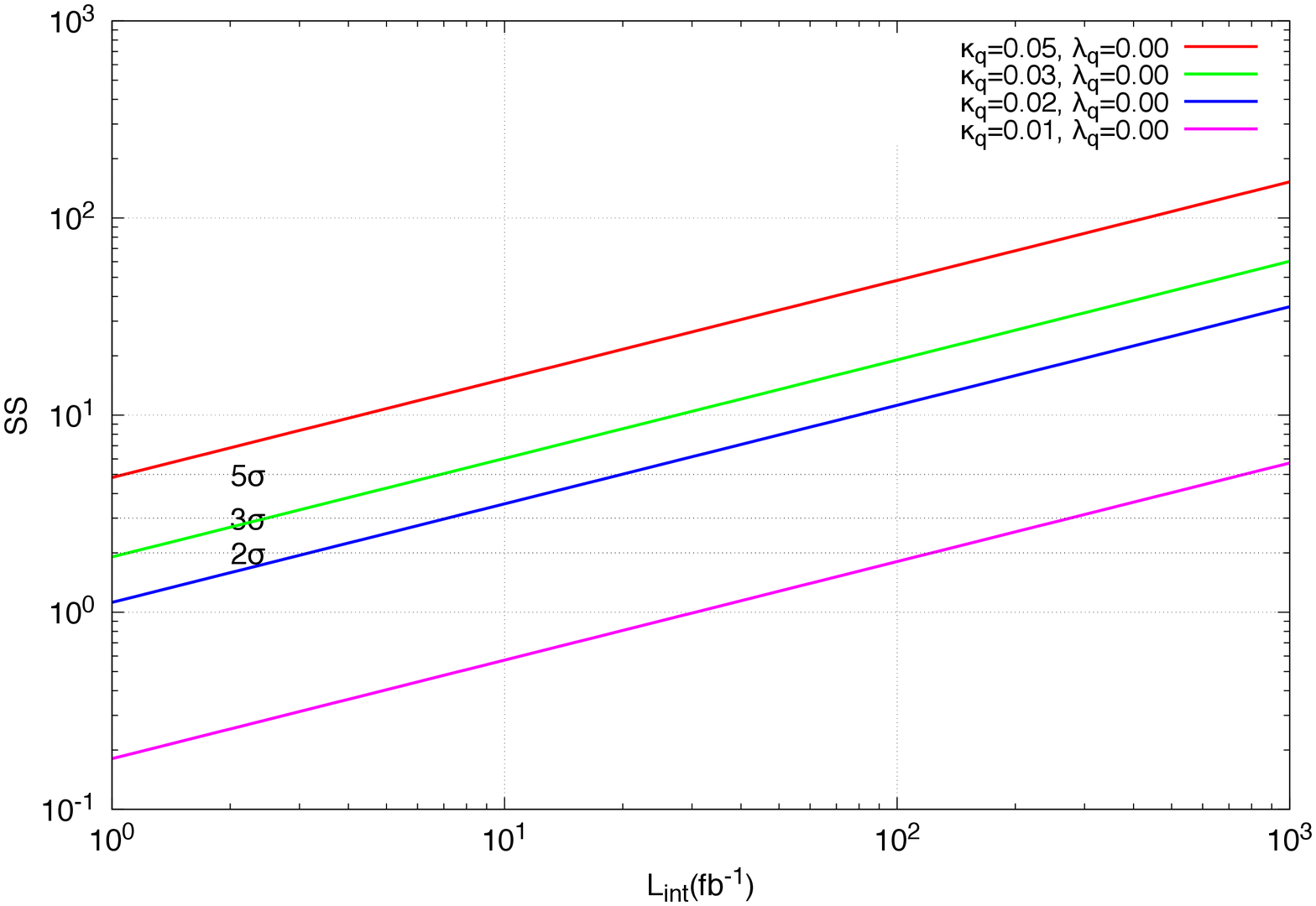}

\includegraphics[scale=0.3]{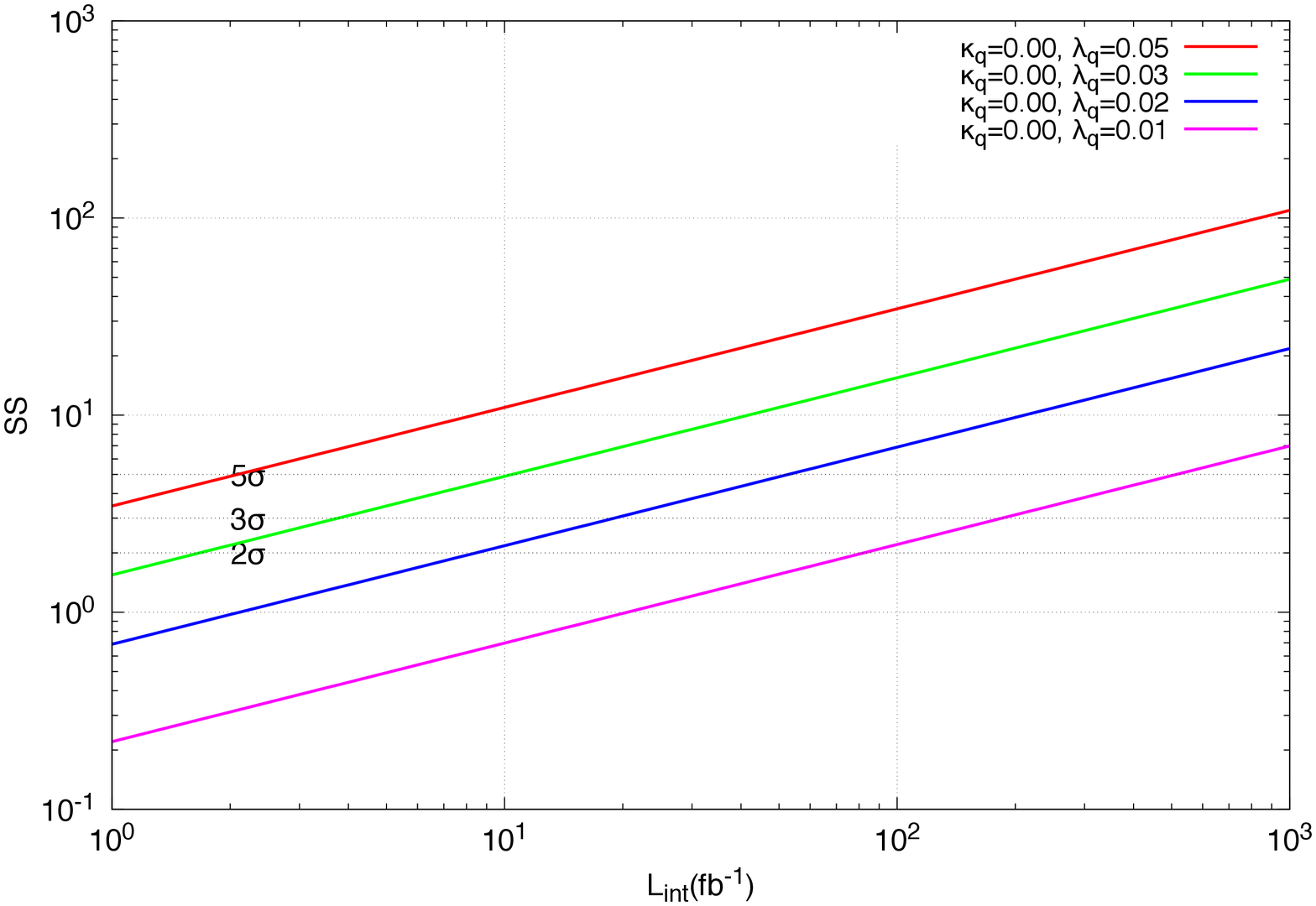}\includegraphics[scale=0.3]{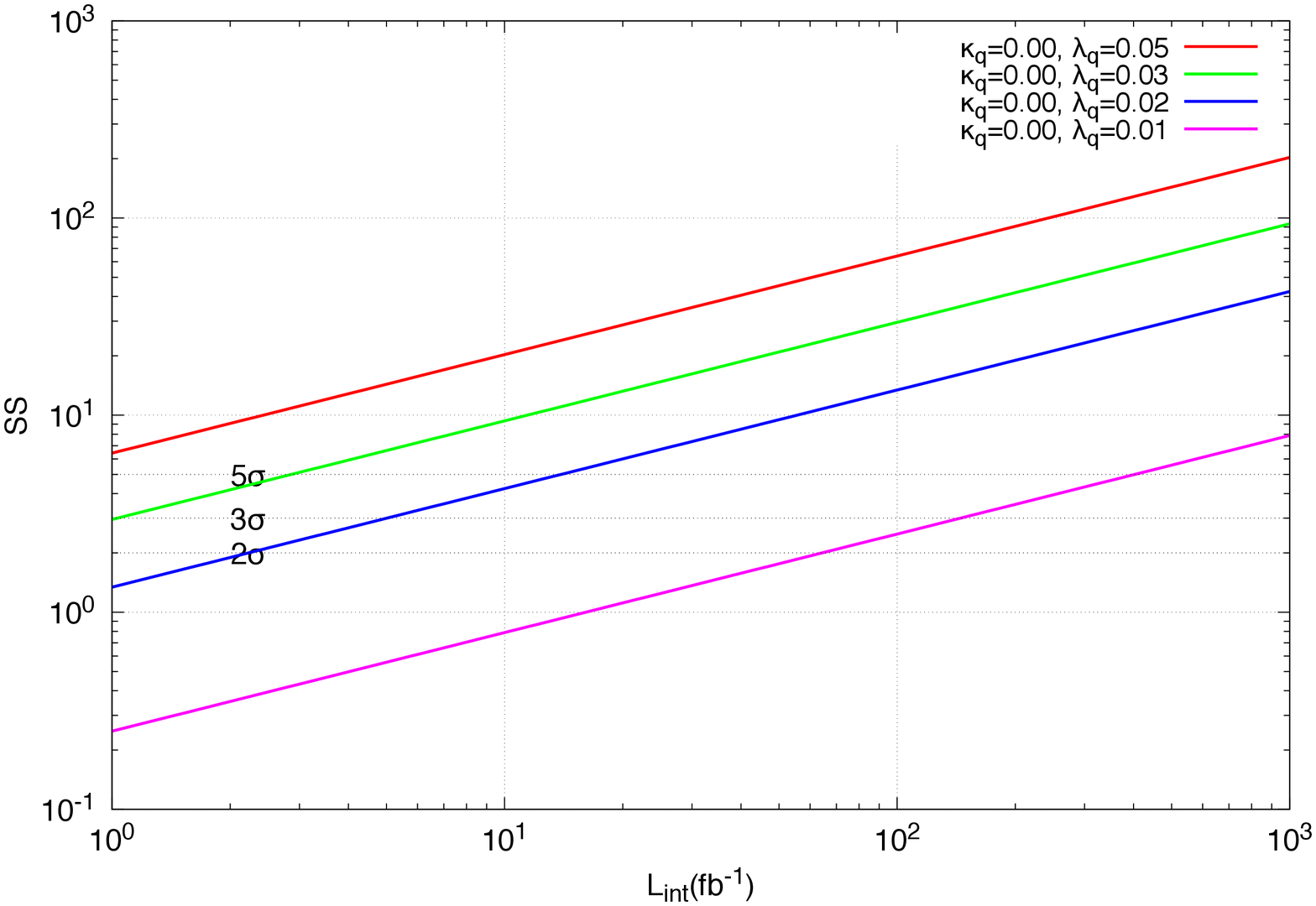}

\includegraphics[scale=0.3]{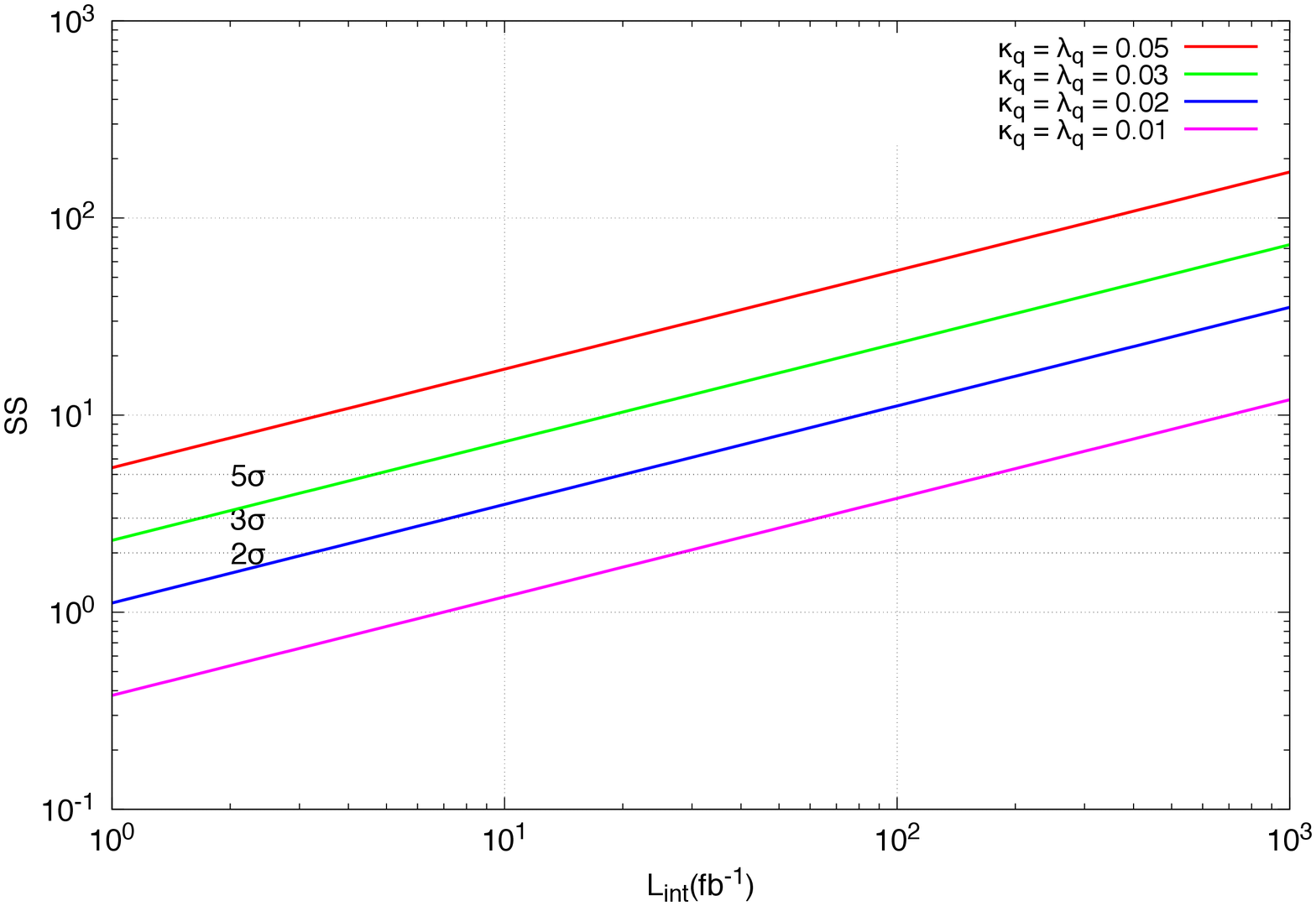} \includegraphics[scale=0.3]{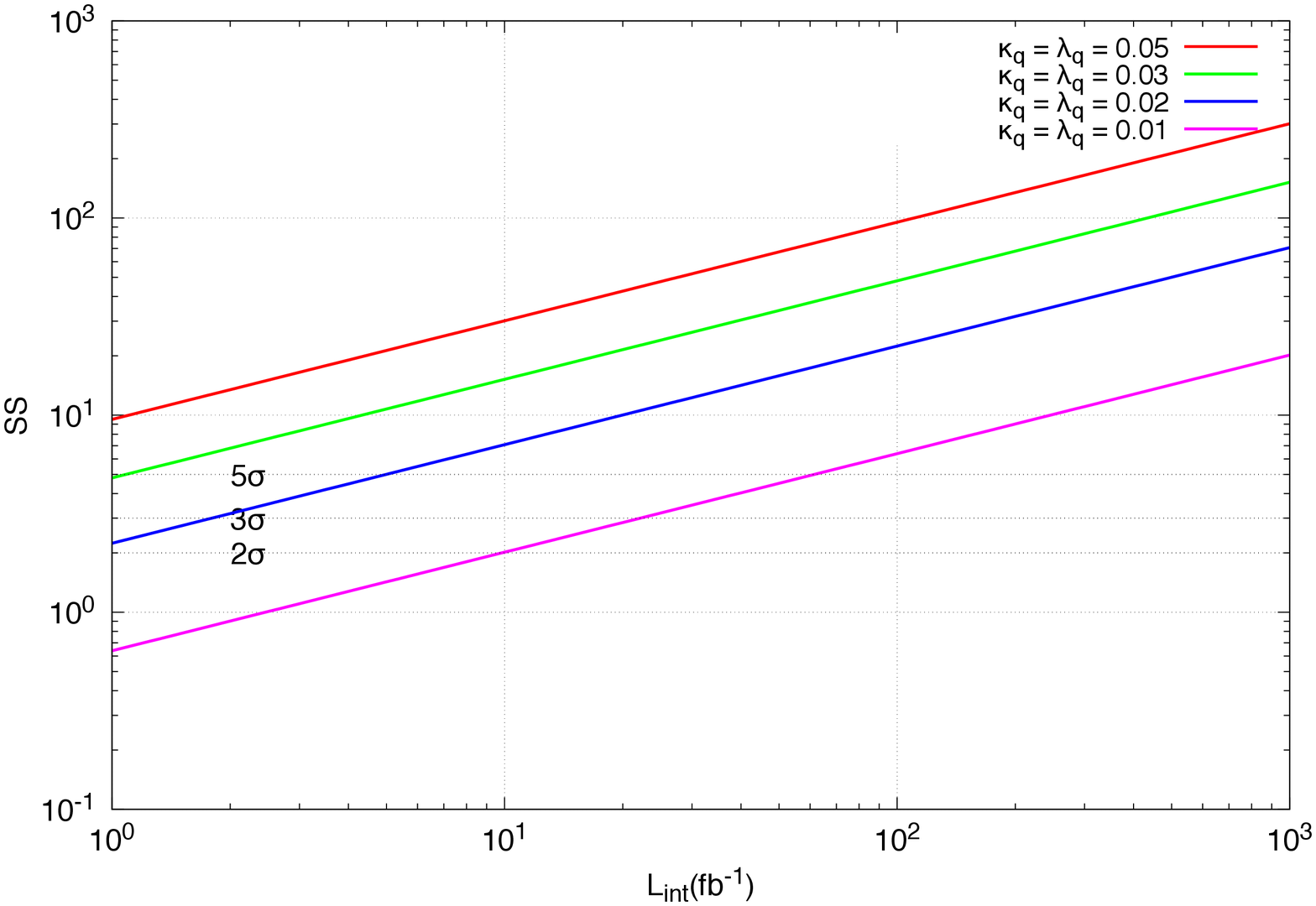}\caption{The statistical significance (SS) for the integrated luminosity ranging
from 100 fb$^{-1}$ to 1 ab$^{-1}$ at the LHeC (first column) and
FCC-he (second column). It includes the contribution from the main
backgrounds on the predicted results. First row shows SS plot for
$\lambda_{q}=0$ while $\kappa_{q}$ changes, second row is for $\kappa_{q}=0$
while $\lambda_{q}$ changes, and third row shows equal coupling scenario
$\kappa_{q}=\lambda_{q}$. \label{fig:fig8}}
\end{figure}

The $SS$ values depending on the integrated luminosity ranging from
1 fb$^{-1}$ to 1 ab$^{-1}$ at the LHeC and FCC-he are presented
in Fig. \ref{fig:fig8} for the coupling scenarios (at first row)
$\lambda_{q}=0.0,\,\kappa_{q}=0.05$, (second row) $\lambda_{q}=0.05,\kappa_{q}=0.0$
and (third row) $\lambda_{q}=0.05,\kappa_{q}=0.05$. The significance
corresponding to $2\sigma$, $3\sigma$ and $5\sigma$ lines (dotted)
are also shown in these figures. In Fig. \ref{fig:fig8}, the $SS$
values depending on the integrated luminosity ranging from $1$ fb$^{-1}$
to $1$ ab$^{-1}$ at the FCC-he are presented for these coupling
scenarios with the $2\sigma$, $3\sigma$ and $5\sigma$ significances.

\begin{figure}[!hbt]
\includegraphics[scale=0.45]{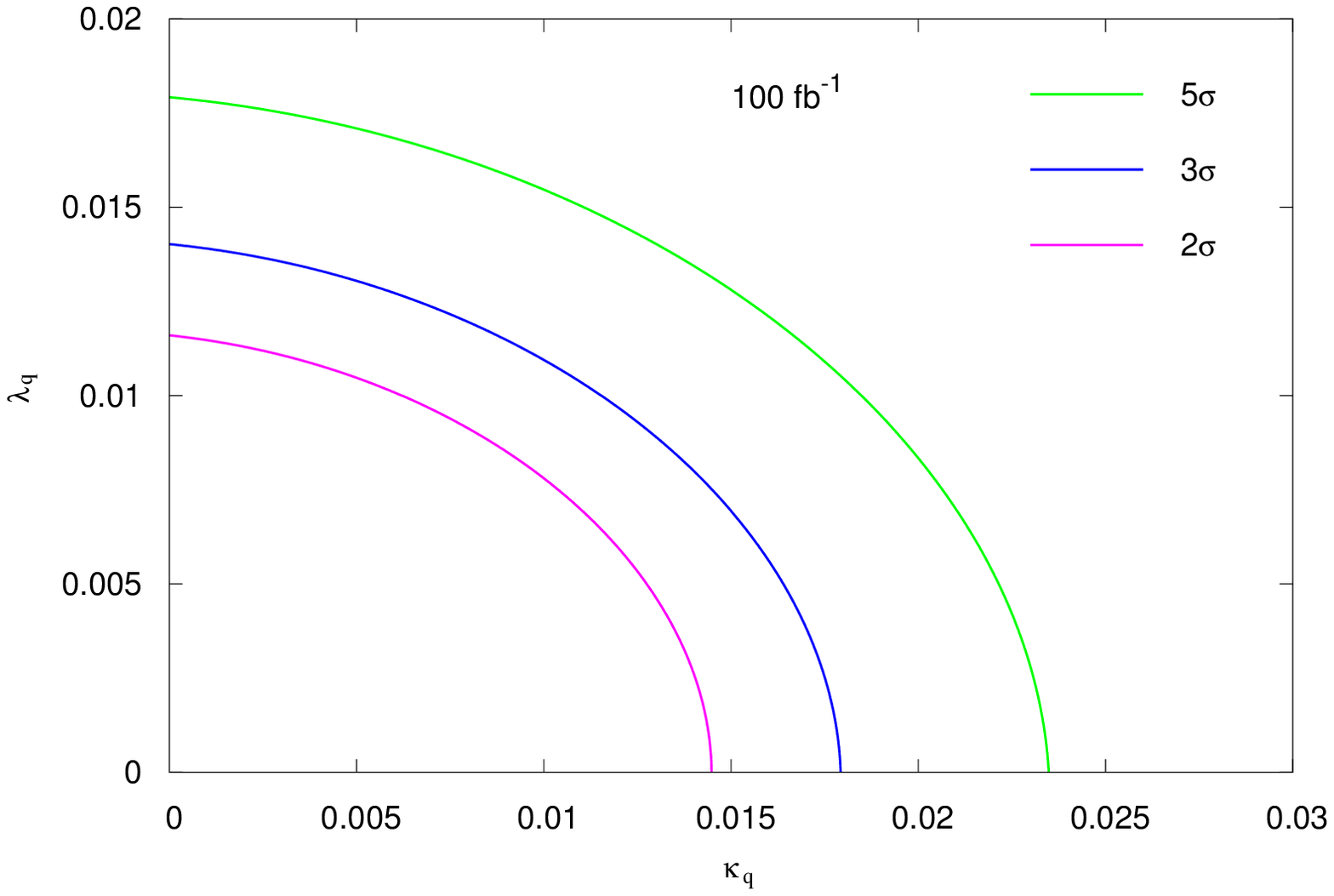} \includegraphics[scale=0.45]{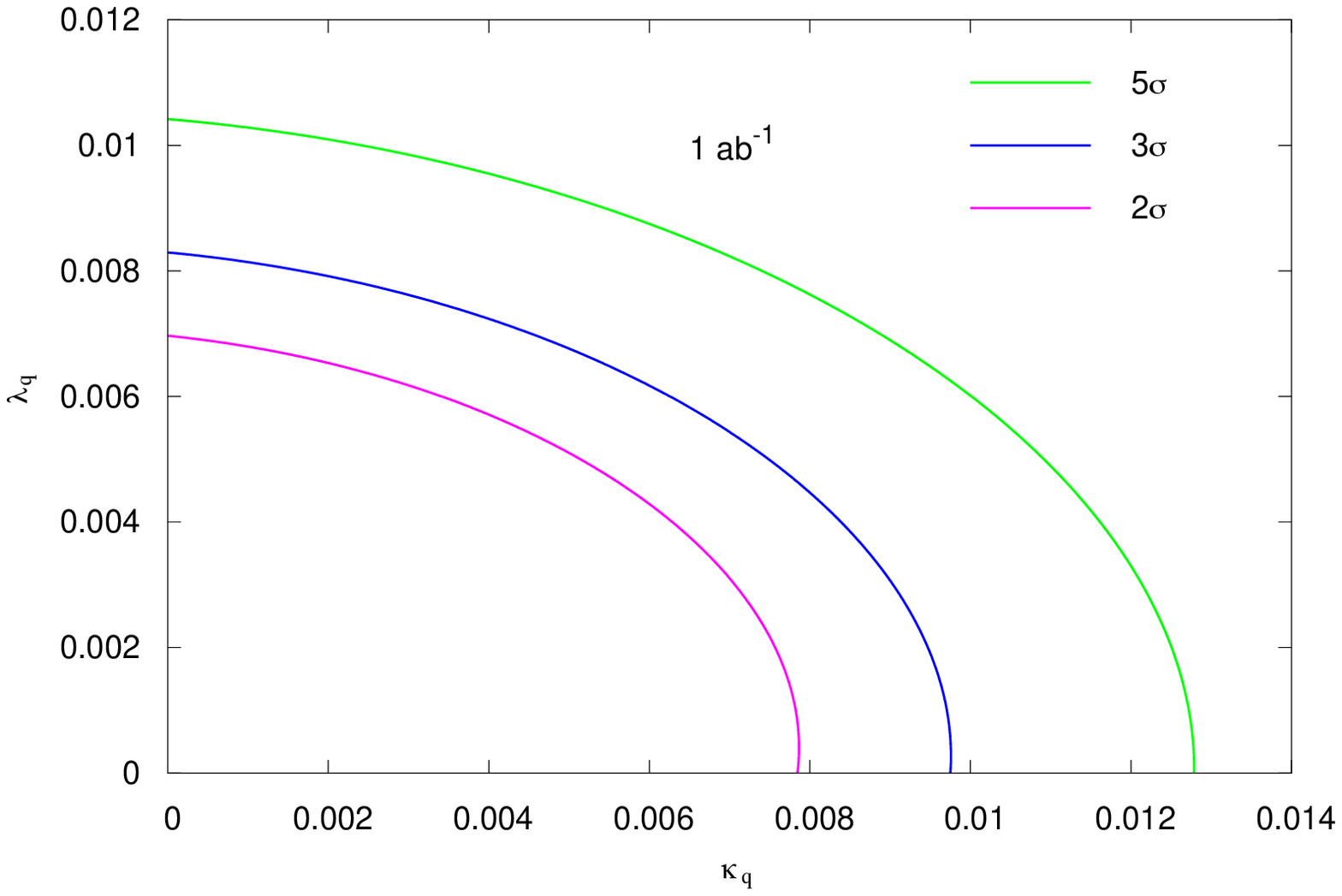}

\includegraphics[clip,scale=0.45]{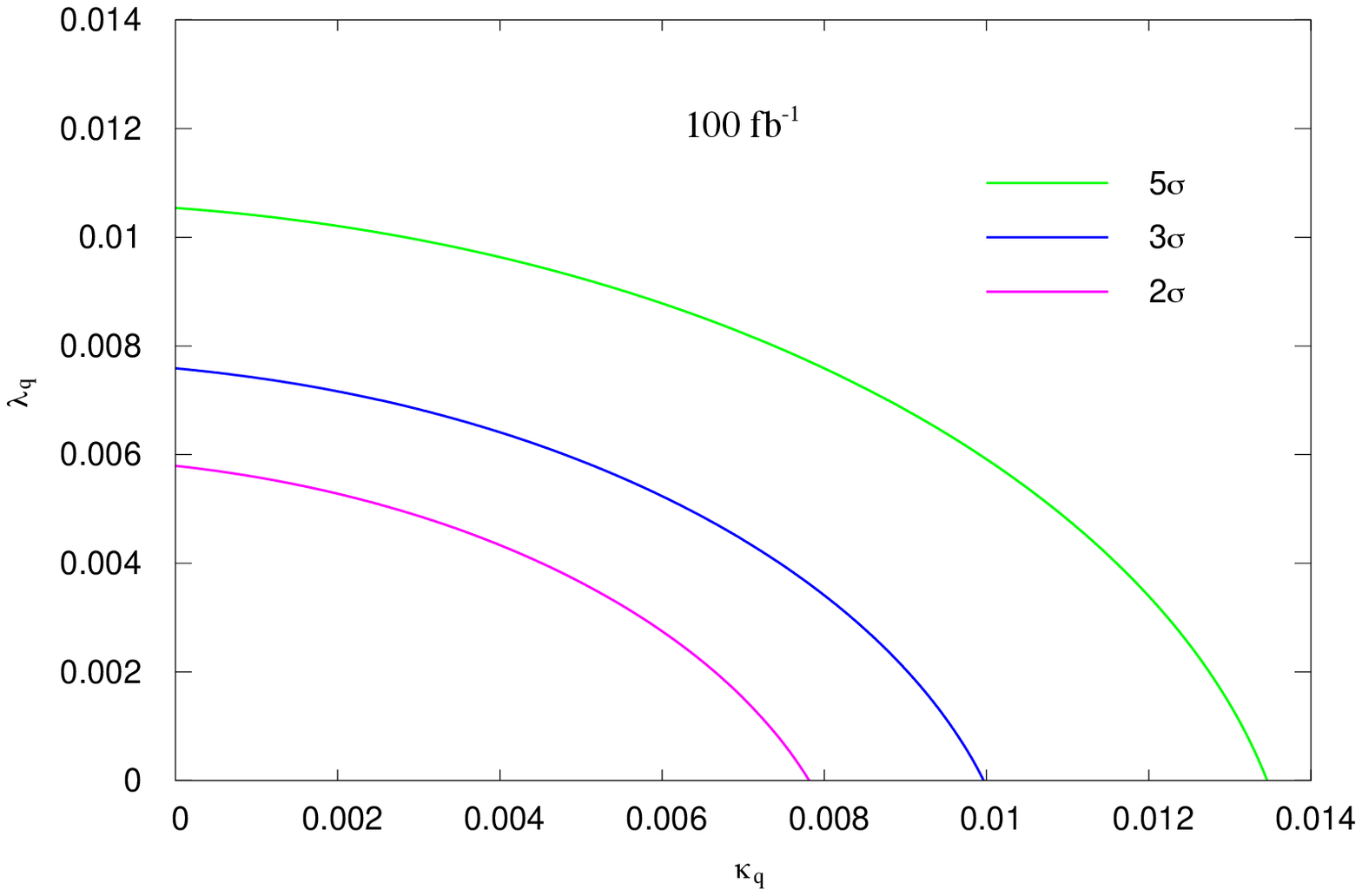}\includegraphics[clip,scale=0.45]{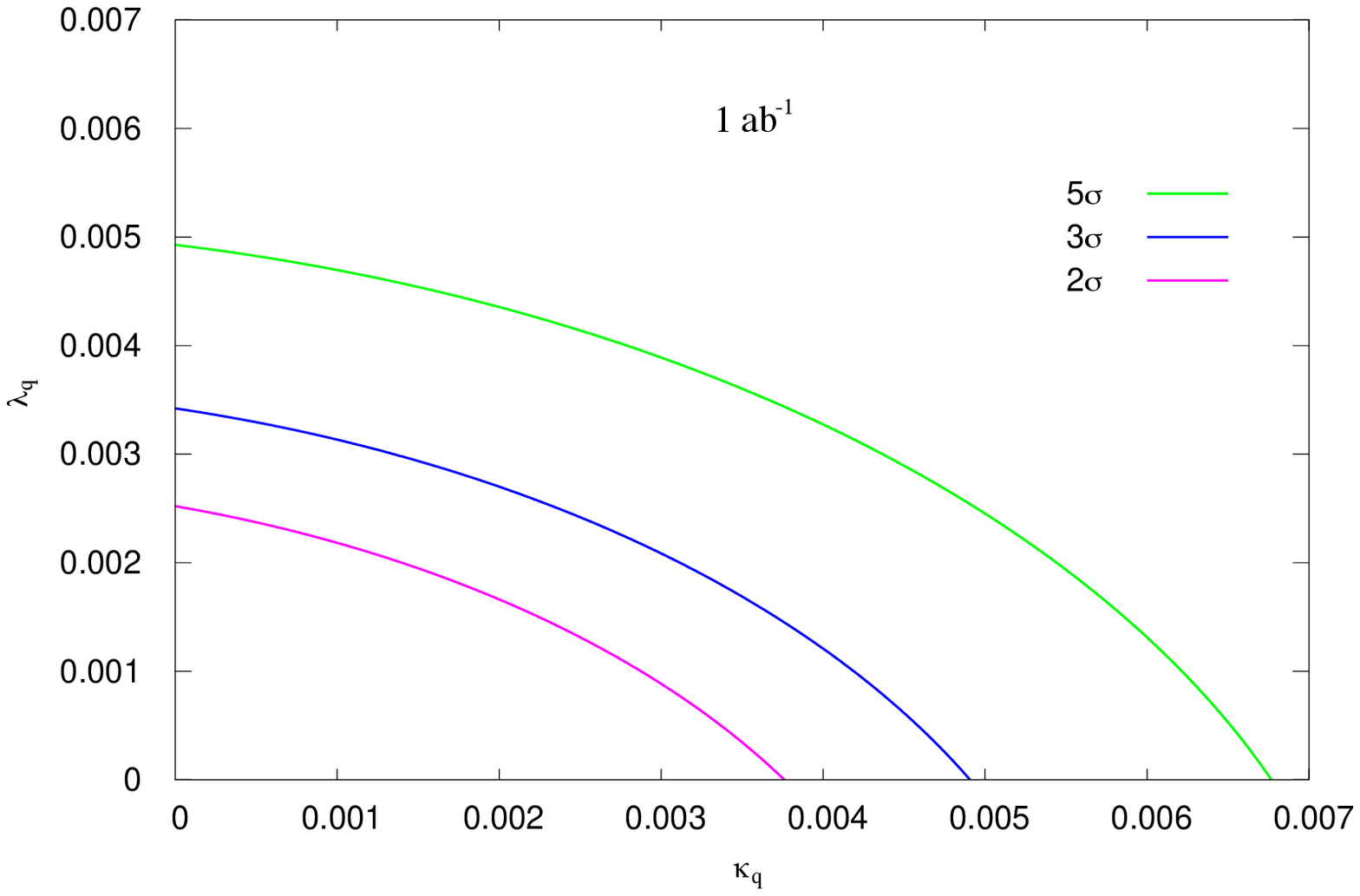}\caption{The reach of proposed couplings at different significance level at
LHeC (first row) and FCC-he (second row) with luminosity projections
100 fb$^{-1}$ and 1 ab$^{-1}$. \label{fig:fig9}}
\end{figure}

Using the corresponding statistical significances, we fit the significance
as a function of two parameters $\kappa_{q}$ and $\lambda_{q}$ at
the integrated luminosity of $100$ fb$^{-1}$ and $1$ ab$^{-1}$.
We obtain contour lines from the fit procedure. In Fig. \ref{fig:fig9},
we estimate the reach for couplings $\kappa_{q}$ and $\lambda_{q}$
corresponding to $2\sigma$, $3\sigma$ and $5\sigma$ significance
for integrated luminosity at the LHeC and FCC-he, respectively. We
obtain the $2\sigma$ significance for the couplings $\kappa_{q}=0.014$,
$\lambda_{q}=0.012$ and $\kappa_{q}=0.008$, $\lambda_{q}=0.007$
at LHeC with the integrated luminosities $100$ fb$^{-1}$ and $1$
ab$^{-1}$, respectively. The sensitivities to the couplings are enhanced
at FCC-he as the obtained values $\kappa_{q}=0.008$, $\lambda_{q}=0.006$
and $\kappa_{q}=0.0037$, $\lambda_{q}=0.0025$ for $L_{int}=100$
fb$^{-1}$ and $1$ ab$^{-1}$, respectively.

The limits on couplings can be translated into the branching ratio
via Fig. \ref{fig:fig1}. We find the upper limits on branching ratio
BR($t\to qZ$)$\leq4.0\times10^{-5}$ and BR($t\to qZ$)$\leq1.0\times10^{-5}$
at $2\sigma$ significance level for $L_{int}=1$ ab$^{-1}$ at LHeC
and FCC-he, respectively. The HL-LHC will produce a large number of
top quarks, which also provide opportunity to search for FCNC processes
to improve existing constraints on the branching ratios BR($t\to qZ$)$<10^{-5}$
with the upgraded LHC experiments. We find better limits when compared
to the current experimental limits and estimations for HL-LHC. In
our previous studies given in Refs. \cite{ICakir2017} and \cite{Denizli17},
we have obtained the limits on the top quark FCNC $tq\gamma$ couplings
depending on the integrated luminosity of future ep colliders. As
a complementary to these studies, here we have analyzed both $tq\gamma$
and $tqZ$ couplings in three different scenarios and obtained sensitivities
to the couplings $\kappa_{q}$ and $\lambda_{q}$.

\begin{figure}
\includegraphics[scale=0.6]{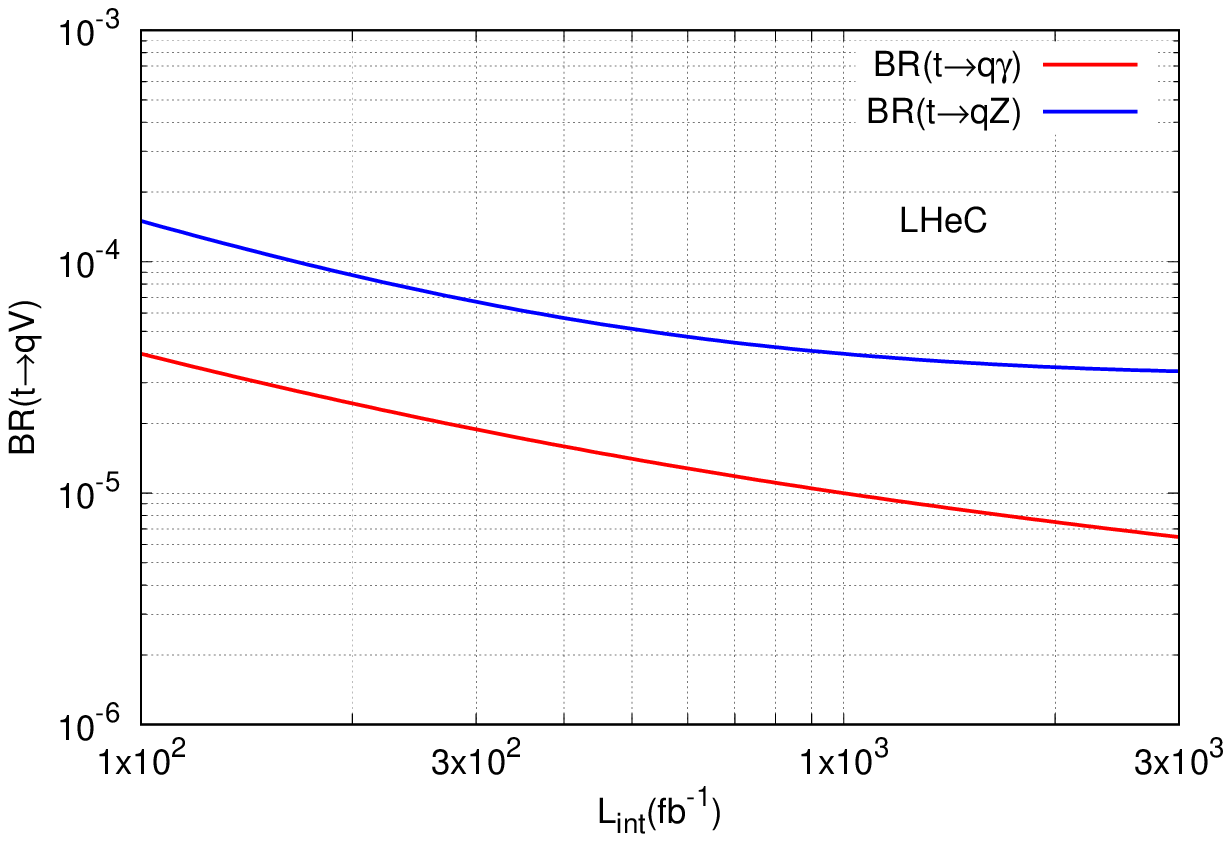}\includegraphics[scale=0.6]{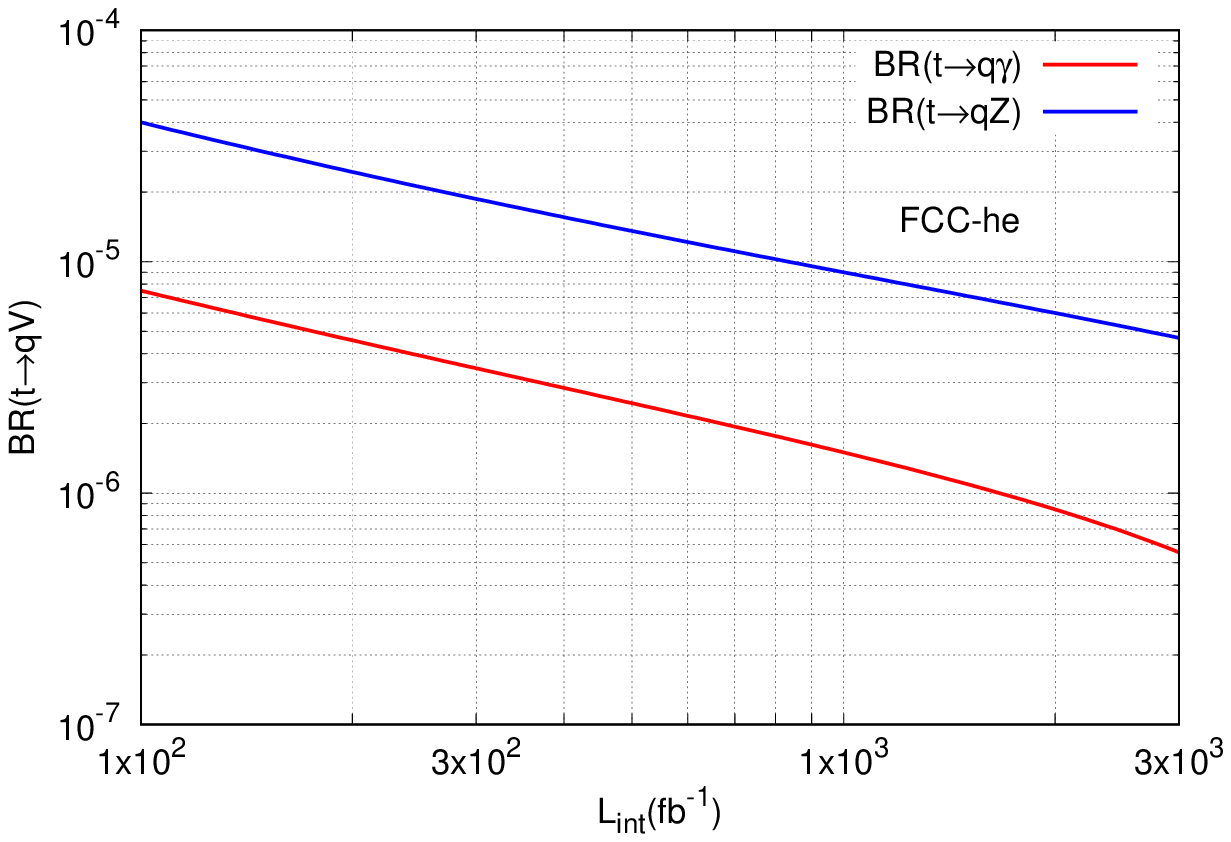}

\caption{Expected sensitivities on BR($t\to q\gamma$) and BR($t\to qZ$) as
a function of the integrated luminosity at LHeC (left) and FCC-he
(right). \label{fig:fig10}}

\end{figure}

Finally, extending the analysis for higher luminosities, we present
the expected sensitivities on BR($t\to q\gamma$) and BR($t\to qZ$)
as a function of the integrated luminosity (in the range between 100
fb$^{-1}$ and 3000 fb$^{-1}$) at the LHeC and FCC-he in Fig. \ref{fig:fig10}.
For the integrated luminosities of 1 ab$^{-1}$, 2 ab$^{-1}$ and
3 ab$^{-1}$, the sensitivities on BR$(t\to q\gamma)$ and BR$(t\to qZ)$
are given in Table \ref{tab:tab4} at the LHeC and FCC-he.

\begin{table}
\caption{The sensitivities to the branching ratios BR($t\to q\gamma$) and
BR($t\to qZ$) for three different luminosity projections at LHeC
and FCC-he. \label{tab:tab4}}

\begin{tabular}{|c|cccccc}
\hline 
Collider &  & \multicolumn{5}{c}{LHeC}\tabularnewline
\hline 
Luminosity &  & 1 ab$^{-1}$ &  & 2 ab$^{-1}$ &  & 3 ab$^{-1}$\tabularnewline
\hline 
BR($t\to q\gamma$) &  & $1.0\times10^{-5}$ &  & $7.5\times10^{-6}$ &  & $6.2\times10^{-6}$\tabularnewline
\hline 
BR($t\to qZ$) &  & $4.0\times10^{-5}$ &  & $3.5\times10^{-5}$ &  & $3.3\times10^{-5}$\tabularnewline
\hline 
\end{tabular}%
\begin{tabular}{|ccccccc|}
\hline 
 &  & \multicolumn{5}{c|}{FCC-he}\tabularnewline
\hline 
 &  & 1 ab$^{-1}$ &  & 2 ab$^{-1}$ &  & 3 ab$^{-1}$\tabularnewline
\hline 
 &  & $1.5\times10^{-6}$ &  & $8.5\times10^{-7}$ &  & $5.5\times10^{-7}$\tabularnewline
\hline 
 &  & $9.5\times10^{-6}$ &  & $6.0\times10^{-6}$ &  & $4.5\times10^{-6}$\tabularnewline
\hline 
\end{tabular}
\end{table}

\section{Conclusion}

The top quark FCNC interactions are important probes for new physics
beyond the SM. It is also worth to mention that the analysis include
the signal and background interference effects. The physics potential
of future $ep$ colliders LHeC and FCC-he for probing new physics
through top FCNC is promoted with their expected complementarity to
the future lepton and hadron colliders. Sensitivities have been achieved
for the $tq\gamma$ and $tqZ$ FCNC couplings at the LHeC with the
center of mass energy of $1.3$ TeV and integrated luminosities of
$L_{int}=1$ ab$^{-1}$, 2 ab$^{-1}$ and 3 ab$^{-1}$. The FCC-he
with higher center of mass energy of $3.5$ TeV will allow us to significantly
improve the sensitivity to the top quark FCNC.

\section{Acknowledgement}

This work was partially supported by Bolu Abant Izzet Baysal University
Scientific Research Projects under the project no: 2018.03.02.1286.
Authors' work was partially supported by Turkish Atomic Energy Authority
(TAEK) under the project grant no. 2018TAEK(CERN)A5.H6.F2-20.

\end{document}